\UseRawInputEncoding
\documentclass[twocolappendix]{emulateapj}

\usepackage{graphicx}% Include figure files
\usepackage{array}
\usepackage{hyperref}
\usepackage{longtable}
\usepackage{amsmath}

\usepackage{siunitx}
\usepackage[outdir=./]{epstopdf}

\usepackage[para]{threeparttable}

\begin{document}

\newcommand{\rotse}    {\sc{ROTSE}}
\newcommand{\rar}       {\rightarrow}
\newcommand{\eps}       {\epsilon}

%\catchline{}{}{}{}{}
%
%%%%%%%%%%%%%%%%%%%%%%%%%%%%%%%%%%%%%%%%%%%%%%%%%%%%%%%%%%%%%%%%%%%%

\title{Cosmological Distance measurement of 12 nearby Supernovae IIP with ROTSE-IIIb}

\author{G. Dhungana\altaffilmark{1,*}}
\author{R. Kehoe\altaffilmark{1}}
\author{R. Staten\altaffilmark{1}}
\author{J. Vinko\altaffilmark{2,3,4,5}}
\author{J. C. Wheeler\altaffilmark{2}}
\author{C. Akerlof\altaffilmark{6}}
\author{D. Doss\altaffilmark{7}}
\author{F. V. Ferrante\altaffilmark{1}}
\author{C. A. Gibson\altaffilmark{7}}
\author{J. Lasker\altaffilmark{1}}
\author{G. H. Marion\altaffilmark{2}}
\author{S. B. Pandey\altaffilmark{8}}
\author{R. M. Quimby\altaffilmark{9,10}}
\author{E. Rykoff\altaffilmark{11}}
\author{D. Smith\altaffilmark{12}}
\author{F. Yuan\altaffilmark{6}}
\author{W. Zheng\altaffilmark{6,11}}

\altaffiltext{1}{Department of Physics, Southern Methodist University, Dallas, TX, USA}
\altaffiltext{2}{Department of Astronomy, University of Texas at Austin, 2515 Speedway, Stop C1400, Austin, Texas 78712-1205}
\altaffiltext{3}{CSFK Konkoly Observatory, Budapest, Konkoly-Thege M. ut 15-17, 1121, Hungary}
\altaffiltext{4}{Department of Optics and Quantum Electronics, University of Szeged, Dom ter 9, Szeged, 6720 Hungary}
\altaffiltext{5}{ELTE E\"otv\"os Lor\'and University, Institute of Physics, P\'azm\'any 
P\'eter s\'t\'any 1/A, Budapest, 1117 Hungary}
\altaffiltext{6}{Department of Physics, University of Michigan, Ann Arbor, MI, USA}
\altaffiltext{7}{McDonald Observatory, University of Texas at Austin, TX, USA}
\altaffiltext{8}{Aryabhatta Research Institute of Observational Sciences (ARIES), Manora Peak, Nainital, Uttarakhand, India, 263001}
\altaffiltext{9}{Department of Astronomy/Mount Laguna Observatory, San Diego State University, 5500 Campanile Drive, San Diego, CA, USA}
\altaffiltext{10}{Kavli Institute for the Physics and Mathematics of the Universe (WPI), The University of Tokyo Institutes for Advanced Study, The University of Tokyo, Kashiwa, Chiba 277-8583, Japan}
\altaffiltext{11}{Department of Astronomy, University of California, Berkeley, California, USA}
\altaffiltext{12}{Guilford College, USA}
\altaffiltext{*}{gdhungana@smu.edu}

\shorttitle{IIP~EPM}
\shortauthors{Dhungana et al.}

\begin{abstract}

 We present cosmological analysis of 12 nearby ($z<0.06$) Type IIP supernovae (SNe IIP) observed with the ROTSE-IIIb telescope. To achieve precise photometry, we present a new image differencing technique that is implemented for the first time on the ROTSE SN photometry pipeline. With this method, we find up to a 20\% increase in the detection efficiency and significant reduction in residual RMS scatter of the SN lightcurves when compared to the previous pipeline performance. We use the published optical spectra and broadband photometry of well studied SNe IIP to establish temporal models for ejecta velocity and photospheric temperature evolution for our SNe IIP population. This study yields measurements that are competitive to other methods even when the data are limited to a single epoch during the photospheric phase of SNe IIP. Using the fully reduced ROTSE photometry and optical spectra, we apply these models to the respective photometric epochs for each SN in the ROTSE IIP sample. This facilitates the use of the Expanding Photosphere Method (EPM) to obtain distance estimates to their respective host galaxies.  We then perform cosmological parameter fitting using these EPM distances from which we measure the Hubble constant to be $72.9^{+5.7}_{-4.3}~{\rm kms^{-1}~Mpc^{-1}}$, which is consistent with the standard $\Lambda CDM$ model values derived using other independent techniques. 

\end{abstract}

\maketitle
\newpage
\section{Introduction} \label{intro}
Supernova (SN) cosmology has matured over the past few decades. 
Supernovae (SNe) have proven excellent distance indicators for astronomy and 
cosmology due to their enormous and standardizable intrinsic brightness. 
Specifically, the improvements in the methods to precisely calibrate distances 
using the luminosities of SNe Ia remain in the forefront of SN cosmology. 
With the arrival of deeper surveys, these methods are being tested at even higher redshifts, and will 
prove complementary to other high redshift cosmological probes such as baryon acoustic 
oscillations (BAO), cosmic microwave background (CMB), and weak lensing (see \cite{weinberg13, nicola17} for a review). 
While SNe Ia yield remarkable precision for the distance calibration based on their peak luminosity and light 
curve width relation (e.g. \cite{phillips93, tripp98, riess98, perlmutter99, des18, brout22}), there remain questions about 
potential impact of discrepancies between the SN Ia models and the actual physical processes occuring \citep{Benetti04, Howell11, marion16, blondin17}. 

SNe IIP are continuously gaining interest as a standardizable candle population
that provides a potent alternative class of distance indicators. 
The SNe IIP are believed to arise from the catastrophic gravitational collapse of the iron core 
of massive stars that have retained a substantial hydrogen envelope even at the time 
of collapse (e.g. \cite{Branchwheeler17} Ch. 12). Because the explosion mechanism as 
well as the radiative transfer in SNe IIP are believed to be better understood than SNe Ia, 
the distance estimation is expected to be less affected by the systematic uncertainties due to explosion physics (e.g. \cite{eastman96}). 
While SNe Ia may exhibit higher absolute luminosity, SNe IIP  explosions also offer tremendous luminosity and are more frequently occurring than SNe Ia (e.g. \citep{graur15}), 
presenting themselves as competitive distance indicators over an extensive redshift baseline. 
While these SNe show substantial diversity in their photometric and spectroscopic 
properties \citep{Filippenko97, hatano99, Faran14, dhungana16, valenti16, Branchwheeler17}, 
strong correlations in photometric and spectroscopic observables during the recombination 
phase can be exploited to provide a distance calibration. Several methods have been 
proposed over the last few decades to make the SNe IIP distance measurements calibratable. 
These methods are generally driven by the correlations of the luminosity with the 
expansion velocities. The pioneering work of \cite{kishner74} using the 
Expanding Photosphere Method (EPM) treated SN IIP as a homologously expanding photosphere that
emits light as a blackbody diluted from atmospheric scattering. 
This method relies on both the photometry and spectroscopy, where the observed 
flux is compared to the effective blackbody flux in the SN rest frame during the photospheric 
expansion phase of the SNe IIP. Using the models for the dilution correction factor (e.g. \cite{dessart05, eastman96}), 
the EPM technique has been applied to numerous SNe IIP from independent 
samples (e.g \cite{schmidt94, hamuy01, jones09, vinko12, bose14, dhungana16, gall16}). 
A close variant based on the correlations of luminosity with the expansion velocity at 50d after explosion 
was suggested and used as the Standardized Candle Method (SCM) (e.g \cite{hamuy02, nugent06, poznanski10, andrea10, deJaeger17, gall18, vogl19, vdyk19, szalai19, dong21}). 
The method has been further generalized (e.g, \cite{kasen09}) to all epochs in the photospheric phase of the events. 
A newer technique called the Photospheric Magnitude Method (e.g \cite{rodriguez14}) is based on empirical color based calibrations 
for the distance. \cite{rodriguez18} used the PMM technique in the near-IR bands where the effects from the dust and 
line contamination is much smaller compared to the optical wavelengths. \cite{deJaeger15} suggested a 
purely photometric technique called Photometric Color Method (PCM) that requires no spectroscopy unlike previous techniques. 

It is important to test 
and improve these methods as we discover more SNe IIP at higher redshifts. These studies involving limited samples of SNe IIP 
at lower redshifts show promising signs of them providing independent and competitive distance estimates. 

We present a cosmological analysis using distances of 12 SNe IIP that were observed by the ROTSE-III telescopes 
during the 2004-2013 survey period. Distances are derived using the EPM technique and primarily based only 
on ROTSE photometry and coordinated optical spectroscopy. This paper is organized as follows. In Section \ref{sec:data}, we describe the 
photometric and spectroscopic data obtained for our SNe IIP sample. Section \ref{sec:reduc} describes the data 
reduction based on a new, improved image differencing technique, along with photometric calibrations, 
spectroscopy and $k$-corrections. In Section \ref{sec:epm}, we summarize the mathematical framework for the EPM. Section \ref{sec:prop} 
discusses the photometric and spectroscopic parameters for the EPM and establishes their time evolution models. 
The EPM distance measurements are discussed in Section \ref{sec:dist}. 
Section \ref{sec:cosmo} presents the cosmological analysis and the Hubble diagram for our SNe IIP sample. We present the results and discussion in Section \label{sec:discuss} and finally 
our conclusions from the paper in Section \ref{sec:conclusion}.
 
\section{Observations} \label{sec:data}
\subsection{Photometry}

Photometric observations were obtained by the ROTSE-IIIb telescope at McDonald Observatory  \citep{akerlof03}. 
The ROTSE-III instruments are 0.45 m robotic Cassegrain telescopes with a $\ang{1.85} \times \ang{1.85}$ field of view (FOV). 
They operate with an unfiltered 2k $\times$ 2k pixel back-illuminated CCD with broad transmission over a wavelength 
range of $3,000-10,000$ \r{A}, achieving a typical limiting magnitude of $\sim18$ mag.

A sample of 12 SNe IIP is obtained from the ROTSE Supernova Survey comprised of three SN search programs spanning 
from 2004 to 2013. A summary table of the 12 events with their host galaxies is given in Table \ref{tab:IIpsample}. 
Each event in this sample has multiple photometric measurements between 1 week and 5 weeks after explosion. We will 
discuss below why this time range is suited for the EPM technique using SNe IIP. 
This sample constituted 4 events from Texas Supernova Search (TSS) (\cite{QuimbyPhd}), 5 events from ROTSE 
Supernova Verification Project (RSVP) (\cite{YuanPhd}) and 3 events from Texas Supernova Spectroscopic Survey (TS$^3$) (\cite{DhunganaPhd}). 
The TSS survey involved the northern sky survey using the ROTSE - IIIb telescope with nightly patrol of 
thousands of galaxies in the nearby clusters. The TSS aimed at amassing a small collection of well observed SNe, targeting 
the earliest possible photometric observation and likewise a triggered spectroscopic followup with the nearby Hobby Eberly Telescope (HET). 
The RSVP survey extended the northern sky with more fields using both the ROTSE- IIIb and IIId telescopes, along with the 
southern sky coverage using the IIIa and IIIc telescopes. The TS$^3$ survey continued with the RSVP fields in automated survey mode, 
however, new triggered follow up photometric and spectroscopic observations were also added for the interesting 
events within and outside the existing ROTSE footprint.
For most of the survey fields, the ROTSE observations are scheduled for 
a paired successive one minute exposure imaging 30 minutes apart. This scheduling is repeated 2-3 times separated by $\sim$2 hours. 
The follow-up auxiliary fields are generally scheduled using the same scheme apart from a few interesting events where imaging cadence is increased. 
In the sample of 12 SNe IIP, 6 were discovered by the ROTSE telescopes, 5 others were observed in the regular survey mode and 1 was observed in triggered auxiliary mode.
\begin{table*} 
\begin{center}
\caption{ROTSE IIP sample for the EPM study}
\label{tab:IIpsample}
%\begin{threeparttable}
\resizebox{\textwidth}{!}{\begin{tabular}{lccccccc}
\hline
\hline
SN & Program/ROTSE Field & Host Galaxy & Spectra & $z$ & $E(B-V)_{tot}$ & Adopted $t_0$(MJD) & References \\
\hline
SN 2004gy & TSS/skc1307+2626 & NGP9 F379-0005009 & 1 & $0.02690\pm0.00100$ & $0.0100\pm0.0007$ & $53362.5\pm2.5$ & 1\\
SN 2005ay & TSS/tss1152+4327 & NGC 3938 & 3 & $0.00270\pm0.00001$ & $0.0183\pm0.0002$ & $53452.5\pm4.0$ & 2,3\\
SN 2006bj & TSS/tss1220+0756 & SDSS J122219.09+073725.5 & 1 & $0.03770\pm0.00100$ & $0.0200\pm0.0004$ & $53815.3\pm3.0$ & 4\\
SN 2006bp & TSS/tss1159+5136 & NGC 3953 & 4 & $0.00351\pm0.00001$ & $0.4000\pm0.0100$ & $53833.7\pm2.0$ & 5,6\\
SN 2008bj & RSVP/sks1155+4643 & MCG +08-22-20 & 1 & $0.01896\pm0.00011$ & $0.0260\pm0.1000$ & $54534.0\pm2.5$ & 4\\
SN 2008gd & RSVP/sks0117+1352 & SDSS J012044.48+144139.6 & 1 & $0.059096\pm0.000053$ & $0.2823\pm0.0582$ & $54726.9\pm3.5$ & 4\\
SN 2008in & RSVP/tss1224+0440 & NGC 4303 & 3 & $0.00522\pm0.00001$ & $0.1000\pm0.1000$ & $54825.1\pm2.1$ & 6,7\\
SN 2009dd & RSVP/tss1209+4958 & NGC 4088 & 3 & $0.00252\pm0.00001$ & $0.3670\pm0.0070$ & $54928.1\pm1.3$ & 8\\
PTF10gva & RSVP/tss1225+1112 & SDSS J122355.39+103448.9 & 1 & $0.02753\pm0.00012$ & $0.0263\pm0.0008$ & $55320.3\pm0.9$ & 9\\
SN 2013ab & TS$^3$/vsp1443+0953 & NGC 5669 & 2 & $0.00456\pm0.00001$ & $0.044\pm0.066$ & $56339.5\pm1.0$ & 10\\
SN 2013bu & TS$^3$/skt2237+3425 & NGC 7331 & 1 & $0.002722\pm0.000004$ & $0.078\pm0.0006$ & $56399.3\pm1.0$ & 11\\
SN 2013ej & TS$^3$/rqa0137+1547 & NGC 0628/M74 & 5 & $0.00219\pm0.000003$ & $0.0610\pm0.0010$ & $56496.9\pm0.3$ & 12\\
\hline
\hline
\end{tabular}}
\begin{tablenotes}\footnotesize
References: \item[1] \cite{guillochon17} \item[2] \cite{galyam08} \item[3] \cite{poznanski10} \item[4] \cite{kelly12} \item[5] \cite{Quimby06} \item[6] \cite{bose14} \item[7] \cite{roy11} \item[8] \cite{pejchapierto} \item[9] \cite{khazov16} \item[10] \cite{bose15} \item[11] \cite{valenti16} \item[12] \cite{dhungana16}
\end{tablenotes}
%\end{threeparttable}
\end{center}
\end{table*}

\subsection{Spectroscopy}
When available, spectroscopic observations were obtained by the Hobby Eberly Telescope (HET) at 
McDonald Observatory \citep{hill98}. The HET possesses a 9.2 m aperture with a 4 arcmin FOV, 
using a $3072 \times 1024$ pixel CCD. The Low Resolution Spectrograph (LRS) is a high throughput optical ($\sim4,200 -- 10,100$ \AA) grism 
spectrograph attached to the HET tracker, with resolving power of $ R=\frac{\lambda}{\Delta\lambda} $ ranging from 600 to 3,000.

Spectra for the SNe IIP sample that were obtained by the HET are archived in the WISeREP(\cite{yaron12}) catalog. For the events 
for which no HET spectrum is available, we obtain them from the literature, WISeREP or Open Supernova Catalog (\cite{guillochon17}). 
 
\section{Data Reduction} \label{sec:reduc}
\subsection{Photometry}
ROTSE III photometry is carried out using standard techniques (\citep{Yuan08, dhungana16}). The online SN pipeline tasked with prompt analysis and SN discovery 
utilizes the image differencing software developed for the RSVP (\cite{Yuan08}). While this differencing is 
robust for the cases where the SN lies substantially out of the host core, 
we have sought to improve the detection efficiency and the root mean square (RMS) scatter when the observation 
is photon limited or close to the host core. Therefore, we developed a new image differencing software 
which we utilized in the offline ROTSE SN photometric data reduction pipeline.

\subsubsection{Image Differencing Technique: Kernel convolution} \label{sec:imagediff}
Image differencing is a common technique used to monitor and characterize the time domain variability of astronomical objects. 
Because of the variation of observational components across exposures, the exact nature of 
variability from the astrophysical source requires proper extraction by correctly modeling the backgrounds in each exposure  
and matching the point spread functions (PSFs). Precise measurements of variability have been performed 
using differencing technique in various circumstances (\cite{Alupton98, Bramich08, kessler15}). We present a new differencing 
software for the ROTSE SN photometry analysis based on the kernel convolution technique (e.g. \cite{Alupton98, Becker12}). 
Due to the complexity of the bright background host with an extended PSF, 
the subtraction technique can yield photometric artifacts. ROTSE was designed to quickly image the large sky areas, thereby maximizing the sky coverage at the cost of spatial resolution. The pixel size of $\sim 1.5"$ is large on the scale of host galaxy morphology. 
It is also challenging to get an accurate measurement of the signal at the noise limit. Allowing a sufficiently 
large basis of PSF variation, this subtraction software,  $\texttt{ImageDiff}$\footnote{$https://github.com/rotsehub/ImageDiff$}, 
attempts to achieve  better performance on the ROTSE image differencing, not only when the background is complicated but also when the signal is photon limited.

Given a recent survey(science) image $S(x,y)$, a higher signal-to-noise template image $T(x,y)$, usually prepared by stacking several past images, 
and a kernel basis {\it K(u,v)}, the survey image is modeled in a linear combination of kernel convoluted template as
\vspace{-0.05cm}
\begin{equation}
\label{eq:kernel}
S(x,y)~=~ (K\otimes T)(x,y) + \epsilon(x,y)
\vspace{-0.05cm}
\end{equation}
Here, $x,y$ are pixel coordinates and $u,v$ are kernel coordinates, $\epsilon(x,y)$ is the error term. 
The kernel basis set constitutes $i$ basis kernels, i.e, ${\it K(u,v) = \{k_{i}(u,v)}\}$. 
We can write Eq. \ref{eq:kernel} as a linear equation
\begin{equation}
S=\sum_{i}A_{i}c_{i}+\epsilon
\end{equation}
 where, $A_{i} = k_{i}\otimes T$ and $c_{i}$ are the coefficients for the linear combination.
We intend to find these coefficients $c_{i}$ corresponding to each kernel $k_{i}$. Assuming Gaussian errors, 
the maximum likelihood (minimum $\chi^{2}$) solution for the coefficient matrix will be
\begin{equation}
\label{eq:chimin1}
{\bf C=(A^{T}N^{-1}A)^{-1}A^{T}N^{-1}S}
\end{equation}
where {\bf N} is the pixel noise matrix, which is a diagonal matrix as the pixel errors are 
treated as statistically uncorrelated. The inverse of the covariance matrix ${\bf A^{T}N^{-1}A}$ must exist. 
A small prior is added at the level of machine precision to ensure the matrix remains well conditioned.

The residual image after subtraction of the background (hereafter difference image) is then simply given in pixel coordinates by 
\begin{equation}
D(x,y) = S(x,y) - \Big(\sum_{i}A_{i}c_{i}\Big)(x,y)
\end{equation}

\subsubsection{Kernel Types}

We use four different kernel bases,
\begin{enumerate}
\item{Sum of Gaussians basis}:
Gaussian functions multiplied by 2-dimensional polynomials
\begin{equation}
k_{i}(u,v)=e^{-(u^2+v^2)/2\sigma_{n}^2} u^p v^q
\end{equation}
where $i$ runs over all permutation of $n$, $p$, $q$. The polynomial order expansion 
used is $0\leq p+q \leq O_n$. The default choices for 3 Gaussians are $\sigma = [0.7,1.5,3.]$ with $O_n = [4,3,2]$. 
The resulting total number of kernels = $\sum_n (O_n+1)\times (O_n+2)/2 = 31$.

\item{Gauss Hermite polynomial basis}: A Gaussian core is multiplied by Hermite polynomials giving
\begin{equation}
k_{i}(u,v)=e^{-(u^2+v^2)/2\sigma^2} H_{m}(u) H_{n}(v)
\end{equation}
where $H_{n}(x) = (-1)^{n} e^{x^{2}} \frac{d^n}{dx^n} e^{-x^2}$ is the $n^{th}$ order Hermite polynomial.
An obvious merit of using the Gauss-Hermite kernels over symmetric Gaussian Kernels is that the former can model asymmetry because 
of the odd-even nature of the Hermite polynomials. An asymmetric PSF can occur for many different reasons such as atmospheric conditions.

\item{Delta function basis}: This kernel constitutes only delta functions
\begin{equation}
k_{i,j}(u,v)=\delta(u-i) \delta(v-j)
\end{equation}
An $11\times 11$ pixel size kernel has 121 orthonormal, single pixel bases.
The benefit of the delta function is that it is shape independent, so there is no parameter to tune. 
However, this may need regularization to ascertain well conditioning of the model to prevent from overfitting. 
See \cite{Becker12} for an application of delta function kernels.

\item{Principal Component Analysis basis:}
We have also adopted a principal component analysis (PCA) based image differencing.
PCA is a technique of reducing data dimensionality without losing any significant feature of the data. 
The principal components are the eigenvectors of the covariance of the dataset. They are sorted by the 
eigenvalues in decending order, i.e., along the component with the highest eigenvalue (first component), the variance is maximized. 
From the kernel-convolved templates using one of the kernels above, we construct an orthogonal eigen-basis using PCA. So in this case, the basis set is transformed from image space to PCA space.
We use the \texttt{empca} package (\cite{bailey12}) to compute the PCA using the expectation maximization (EM) method. 

\begin{table}
\begin{center}
\caption{$\chi^{2}$ and $R$ values of differencing using different kernels for ROTSE field rqa0137+1547 shown in Fig. \ref{fig:figdiff1}.}
 \label{tab:tabdifference}
\begin{tabular}{lccc}\hline
Kernel Type & $\chi^{2}/dof$ & R value & Pull \\
\hline
Gaussian sum & 0.82 & 0.97 & $\mathcal{N}(-0.01,0.91)$\\
Gauss Hermite & 0.82 & 0.97 & $\mathcal{N}(-0.01,0.90)$ \\
Delta Function & 0.81 & 0.97 & $\mathcal{N}(-0.01,0.90)$ \\
EMPCA & 0.85 & 0.93 & $\mathcal{N}(0.03,0.94)$\\
\hline
\end{tabular}
\end{center}
\end{table}

The top panel of Fig. \ref{fig:figdiff1} shows a high S/N template image, a later survey image and the output 
difference image for a subimage of ROTSE SN field \MakeTextLowercase{rqa$0137+1547$}. The bottom panel shows a slice of the galaxy profile, 
the model PSF using the sum of Gaussian kernels and the residual from the subtraction. The right most plot shows the Pull distribution for the residuals 
in normal form ($\mathcal{N}$) defined as 
\begin{equation}
Pull=\sum_{i}\frac{data_{i}-model_{i}}{\sigma_{i}}
\end{equation}
where $i$ runs over all the pixels in the subimage. A performance summary of the image differencing algorithm using different kernel 
types on the same field survey image is shown in the Table \ref{tab:tabdifference}.

\begin{figure} 
\includegraphics[width=0.48\textwidth]{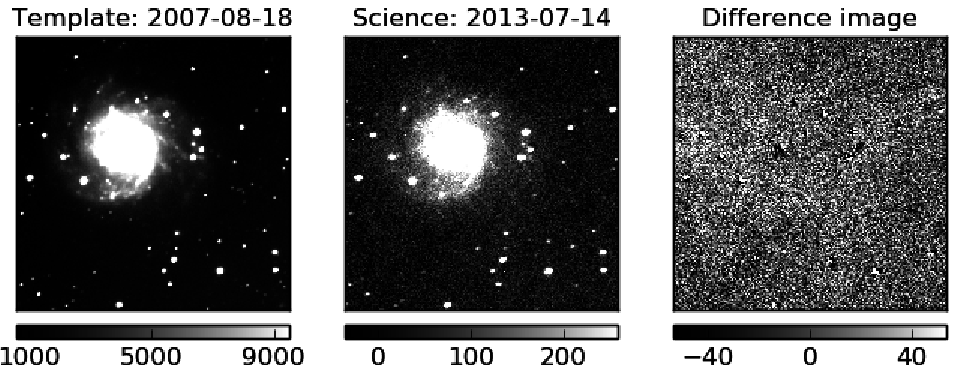}
\includegraphics[width=0.5\textwidth]{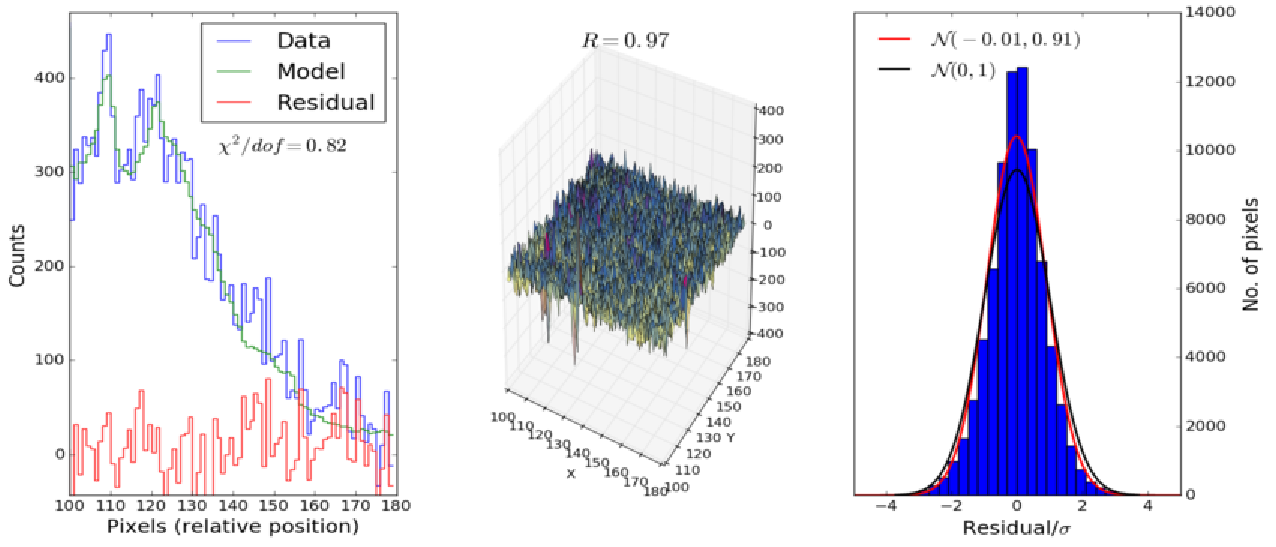}
\caption{Top: A high S/N template subimage (left) and a survey subimage (middle) of the ROTSE IIIb field  rqa$0137+1547$, 
and the difference image using \texttt{ImageDiff} (right). {\it Bottom}: A slice from the center of the subimage 
showing the galaxy $M74$ profile, the fit model and the residual (left); same residual in 2D shown in $80\times 80$ pixels for clarity (middle), 
where R is the measure of the fraction of observation variance preserved in the difference image; pull distribution showing a Gaussian fit (red) yielding 
residual $\mathcal{N}(-0.01,0.91)$ and the theoretical zero mean, 
unit variance standard normal distribution (black) in the right panel. The pull is close to standard normal distribution.}
\label{fig:figdiff1}
\end{figure}
\end{enumerate}

%\paragraph{Performance}
%~\\
\subsubsection{Performance of Image Differencing}
\label{sec:imagediffperf}
\begin{figure}
\begin{center}
\includegraphics[width=0.45\textwidth]{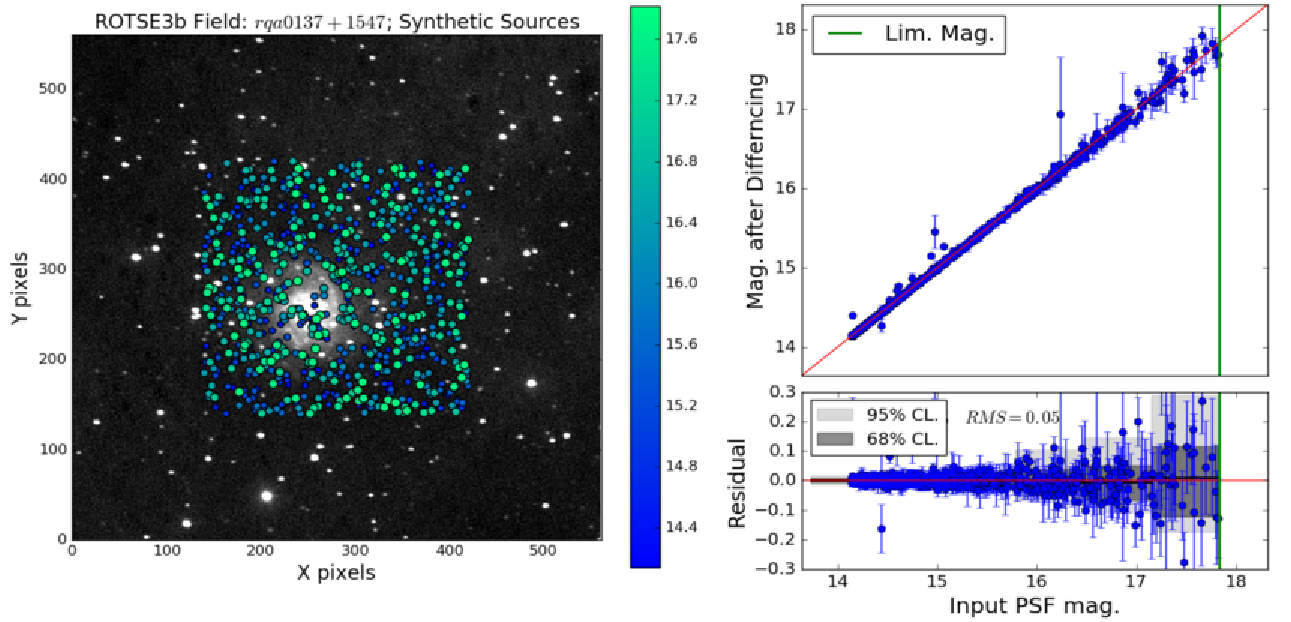}
\caption{Left: 1000 simulated sources superimposed on a data image. Points show injected sources but do not represent the true PSF shape of the simulated objects. Right: Performance of \texttt{ImageDiff}. Overall residual mean is 0 yielding no bias, RMS $\sim$0.05 magnitude; although close to the limiting magnitude, RMS $\sim$0.1 mag. Pull distribution is $\sim \mathcal{N}(0.1,~0.94)$}.
\label{fig:diffsim}
\end{center}
\end{figure}

We monitor the performance of the differencing algorithm for both the spatial and temporal PSF variation in the following two ways. 
We simulate the PSF in order to establish the proper performance of the template subtraction, and we consider lightcurve 
properties to comment on efficiency and stability of source photometry. 
We consider the former using a PSF model profile with a Gaussian core and a wing component, allowing ellipticity variation from \cite{bolton10}: 
\begin{eqnarray}
I(x,y)=\frac{(1-b)}{\sqrt{2\pi}\sigma}e^{[\frac{-r_{ell}^2}{2\sigma^2}]} + \frac{b e^{(-r/r_0)}}{2\pi r r_0} \\
r_{ell} = \sqrt{qx^2+y^2/q}
\end{eqnarray}
where $b$ controls the wing contribution, $r$ is the radial offset from the PSF center, $r_{0}$ is the characteristic size of the wing, 
$q$ is the ellipticity and $x~\&~y$ are related to CCD coordinates by rotation/translation.
Monte Carlo (MC) simulation is performed by injecting objects of random magnitudes at random locations within a subimage of a ROTSE survey image. 
To disallow tight blending of the injected source with the point data sources in the image, a \texttt{scikit} (\cite{pedregosa12}) $k-d$ tree query 
is performed taking a radius of 1 FWHM of the PSFs of the data image, derived by  \texttt{sextractor}\citep{bertin96}. 
For the injected sources, the image subtraction is performed iteratively one by one and the final photometry is performed on the difference image.  
The extracted magnitudes are compared with the input magnitudes. An example simulation for the ROTSE field $rqa0137+1546$ is shown in Fig. \ref{fig:diffsim}. 
The RMS of the photometry residuals is at the 0.05 magnitude level, and a pull distribution shows normal $\sim \mathcal{N}(0.1,~0.94)$ distribution.

\begin{figure}
\includegraphics[width=0.48\textwidth]{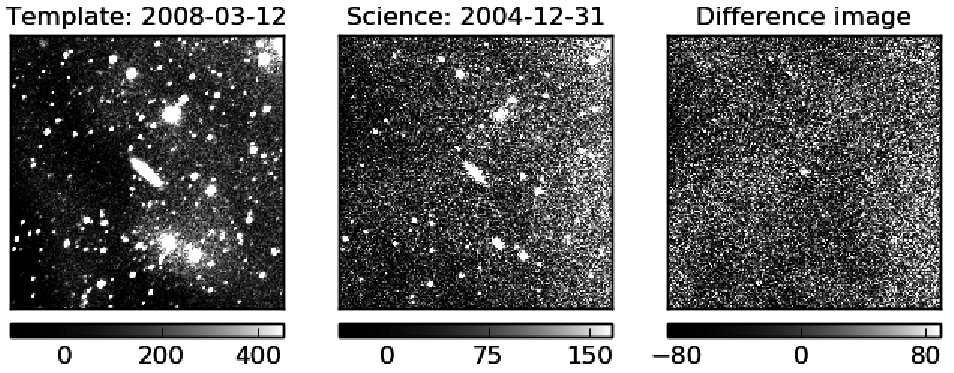}
\includegraphics[width=0.5\textwidth]{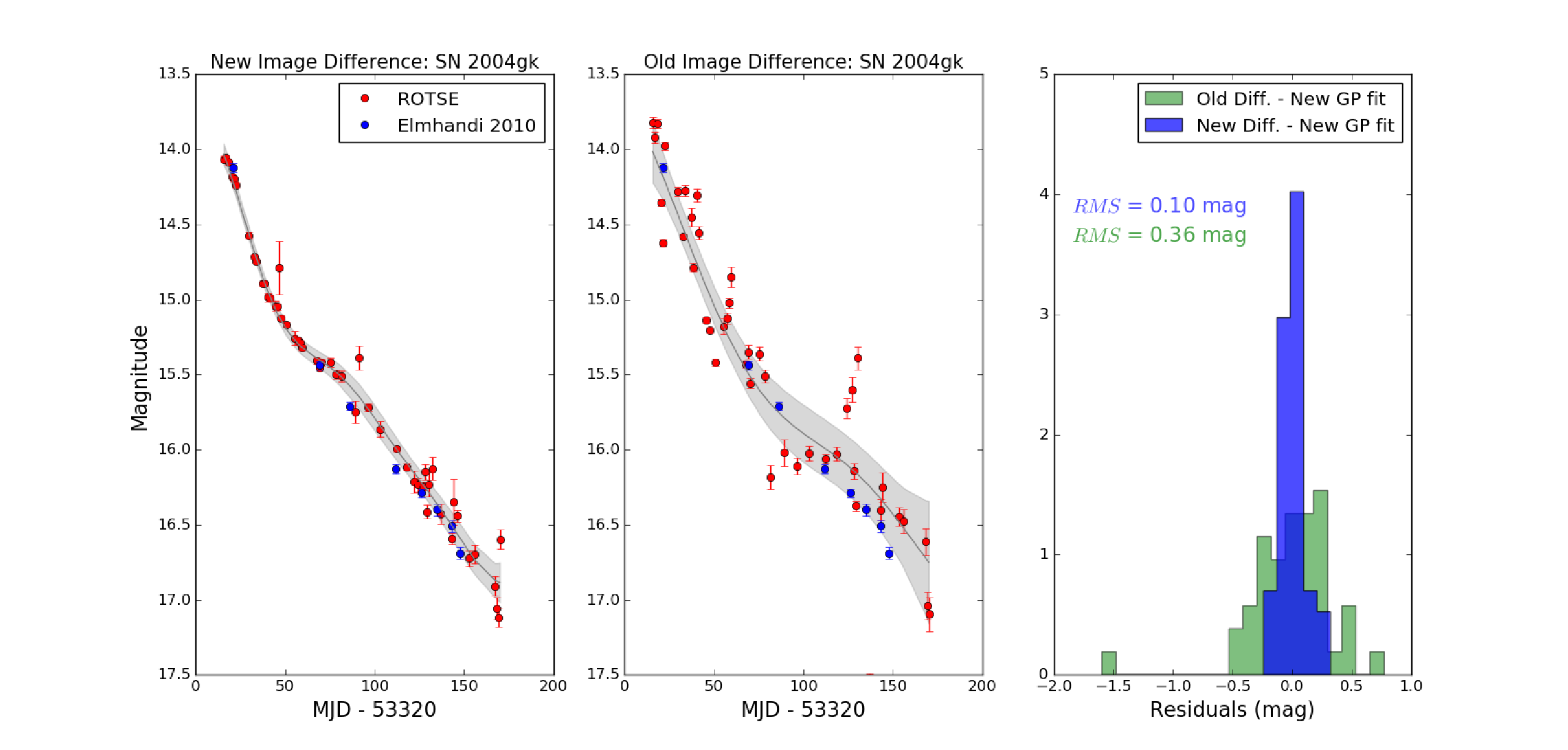}
\caption{Top: Image subtraction of a $280\times 280$ pixel subimage of ROTSE tss1246+1249 field: a high S/N template image (left) obtained with stacking 30 past images, 
a survey image with potential SN at the center(middle) and the difference image from \texttt{ImageDiff} using the sum of Gaussian kernel convolution of template image. SN 2004gk is clearly visible at the center of the difference image.
Bottom: Photometric performance using new and old image differencing methods for SN 2004gk. The reduced data points are normalized to \cite{elmhamdi11} V band magnitude (shown in blue points), on MJD 53389.0 for training the eye. The solid line is a Gaussian Process regression fit, with the filled region being the 95\% confidence posterior prediction. The rightmost panel shows the residuals of old (green) and new (blue) light curves after subtracting the respective best fits. The rms scatter for the green and blue histograms are respectively 0.36 and 0.10 mag, and the pulls on the new image differencing yield a dispersion of 1.03, which is substantially improved compared to the old differencing, where the pull yields a dispersion of 2.82.}
\label{fig:performdiff1}
\end{figure}

In most of the SNe analysed, the new image subtraction yields 10-20\% more detections, and the scatter 
of the residuals also is remarkably narrower. The top panel of Fig. \ref{fig:performdiff1} shows an example of image differencing of an epoch of SN 2004gk using the new image subtraction software. The SN is clearly observerd on the residual image on the right. The bottom panel shows the lightcurve of SN2004gk obtained using the old and new image subtraction. 
Each light curve is fitted with Gaussian Process (GP) regression using \texttt{scikit}. The rightmost plot shows the residuals of new and 
old light curves obtained after subtracting the GP best fit models. It is observed that the new image differencing not only has higher 
detection efficiency but also has over 3.5 times improvement in the residuals scatter. The pull distributions are found to be $\mathcal{N}(0.01,~1.03)$ 
for the new and $\mathcal{N}(0.39,~2.82)$ for the old method; suggesting no significant bias due to the new method. 
The typical pulls on the other SN light curves obtained with \texttt{ImageDiff} also follow within 10\% of a standard normal distribution $\mathcal{N}(0,1)$ . 

%\end{itemize}

\subsubsection{Photometric Calibration}
\begin{figure}
\begin{center}
\includegraphics[width=0.5\textwidth]{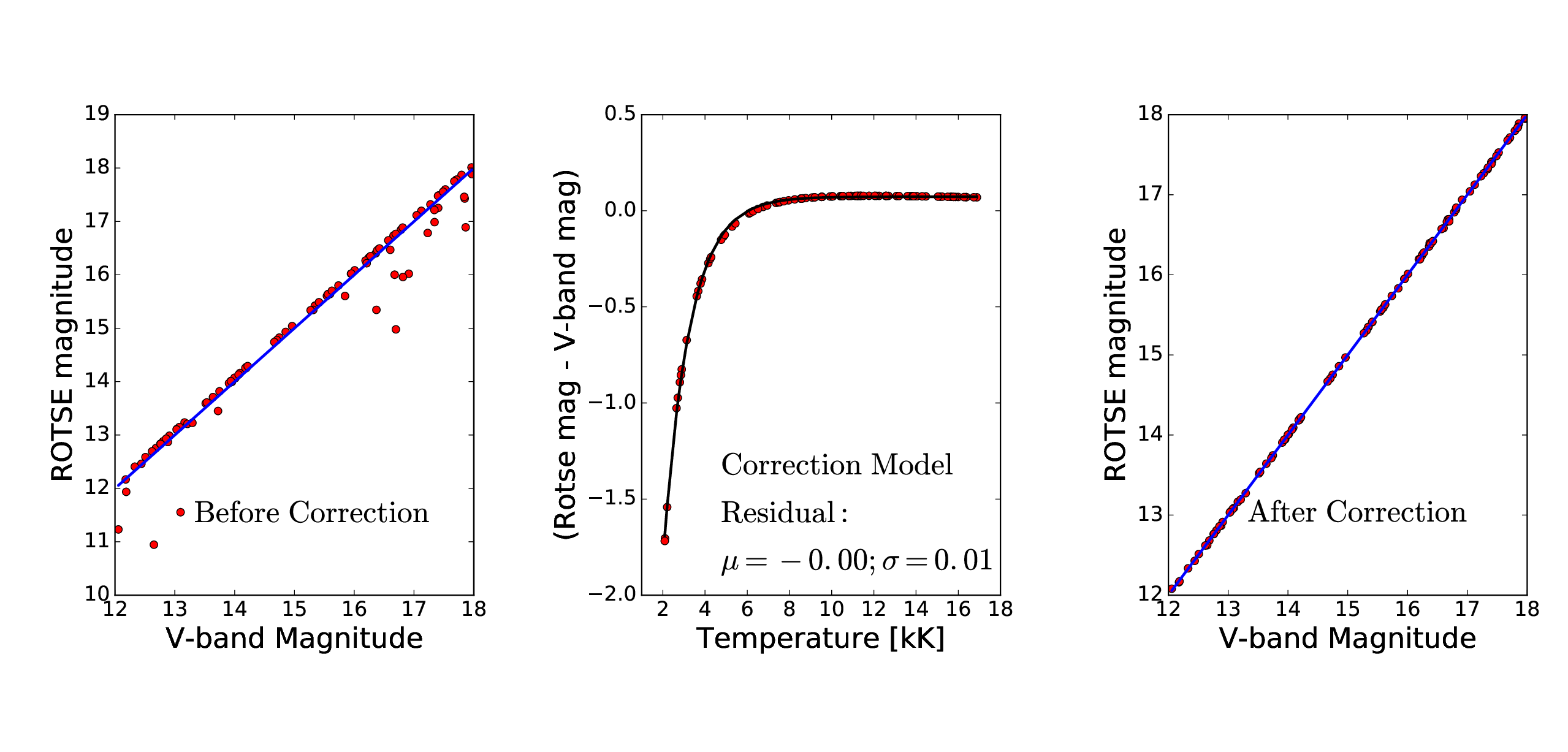}
\caption{Correction of systematic effects of calibrating ROTSE magnitudes to catalog $V$ band data. 
Shown are the offsets in scatter plot for 100 Monte Carlo blackbody spectra spanning 2000K to 17000K temperature, randomly normalized to $V$ 
band mag range 12-18 (left), offsets varying with temperature and the correction model given by Eq. \ref{eq:rotseVcalib}(middle) 
and offsets after applying correction (right).}
\label{fig:rotseVcalib}
\end{center}
\end{figure}

ROTSE magnitudes are calibrated to magnitudes from the  APASS\footnote{$https://www.aavso.org/apass$} DR9 catalog. 
These ROTSE magnitudes are corrected for extinction modeled from \cite{sf11}. Note that the ROTSE CCD response is significantly different 
than the $V$ band filter response function. To establish a concrete and accurate calibration for the rapidly evolving 
Spectral Energy Distribution (SED) of SNe IIP, we quantify any potential offset associated with the calibration of 
ROTSE flux with the $V$ band flux of the field stars. We perform Monte Carlo simulations of blackbody continuum spectra 
of varying temperature 2 kK- 17 kK, randomly normalized to $V$ band magnitude in the range 12-18. The selection of temperature 
range (Section \ref{sec:temp}) and magnitude range agrees with the observed SN IIP temperatures and magnitudes for epochs similar to those in the sample. 
The blackbody spectra are convolved with the ROTSE response function and the $V$ band filter response function, and magnitudes are estimated. 
The left plot of Fig. \ref{fig:rotseVcalib} shows the comparision of the simulated magnitudes for the ROTSE and $V$ band. 
It is clear from the scatter that there is need for correction in both directions. To address these offsets, we use an exponentially growing 
function of temperature ($T$) 

\begin{equation} \label{eq:rotseVcalib}
m_{ROTSE,V} - V = a+b(1-e^{cT})\equiv corr
\end{equation}
where $m_{ROTSE,V}$ is the magnitude obtained from calibrating to $V$ data before correction. The final calibrated magnitude is then $m_{ROTSE,V}-corr$.
Best fit model parameters obtained using the simulation of 100 random blackbody spectra 
yield $a=-9.46\pm0.11$; $b=9.52\pm0.11$ and $c=(8.06\pm0.048)\times10^{-4}$. The state before correction, 
the correction model and magnitudes after applying the correction for the 100 Monte Carlo sample 
are shown in the middle and the right plot of Fig. \ref{fig:rotseVcalib}. 
The RMS in the residuals is about 0.01 mag and no fundamental bias is observed from the correction. The residual RMS is much smaller than the typical 
statistical uncertainty of the obtained ROTSE magnitudes and is propagated as uncorrelated systematic uncertainty in the final photometry. 
As broadband observations are not available for the events in the sample at all epochs; and we are explicitly measuring and modeling temperature evolution in Section \ref{sec:prop}, 
we used a temperature dependent correction model. A color dependent correction would require broadband observations at the photometric epochs. 
When compared with the available $V$ band data of several IIP SNe in the sample, the corrected photometric measurements during 
the plateau phase are obtained to be statistically consistent.

\subsection{Spectroscopy}
HET data are reduced using standard spectroscopic techniques as described in (\cite{silverman12, dhungana16}). 
Other spectra are obtained in the fully reduced form from the literature or from databases. All the spectra are converted 
to SN rest frame and corrected for reddening using the \cite{Fitzpatrick99} model.  \\

\subsection{Extinction, redshift and explosion epochs}\label{sec:exporp}
Table \ref{tab:IIpsample} provides our adopted extinction, redshift and the explosion epochs for each event in our sample. 
We refer to the literature for respective measurements. When no extinction is available for the host, we adopt MW extinction 
only using \cite{sf11}. When the explosion epochs are not available, we adopt the arithmetic mean of the first detection and 
latest non detection in the ROTSE photometry or such reported in the literature, whichever provides the best constraint. 
For our sample, explosion epochs have uncertainty of 0.3 to 4 days. 
The redshifts are taken from the literature or the galactic redshift from the NED\footnote{https://ned.ipac.caltech.edu/} database unless otherwise noted.

\subsection{K-Correction}
\begin{figure}
\begin{center}
\includegraphics[width=0.45\textwidth]{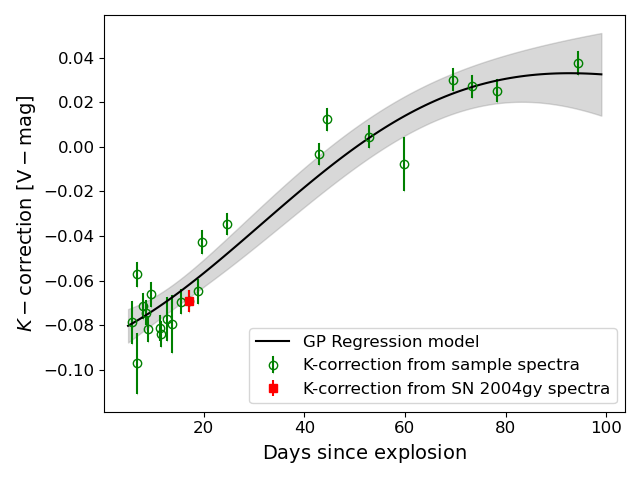}
\includegraphics[width=0.45\textwidth]{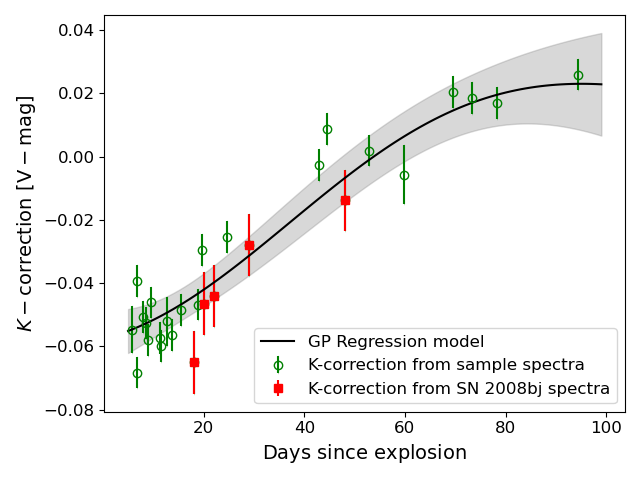}
\caption{Top: $K$-corrections derived spectroscopically from the $z<0.01$ SNe sample (green circles) and the SN2004gy spectrum (red squares) for $z=0.0269$, 
corresponding to SN 2004gy. The solid line is the best fit GP model and the shaded region represents the 68\% confidence region for the posterior. 
Bottom: The same as top for $z=0.019$, corresponding to SN 2008bj.}
\label{fig:Kcorrection}
\end{center}
\end{figure}
Since we do not have color information from broadband photometry for all the events in our sample, we perform a spectrophotometric 
approach to obtain the $K$-correction for events with $z>0.01$. First, to determine the $K$ correction for each event, the spectroscopic 
data of all the nearby SNe ($z<0.01$) samples during the plateau phase are redshifted by the value for the SN considered and 
and both the observer frame and the SN rest frame magnitudes are taken by applying the $V$ band filter. 
The difference of the rest frame magnitudes from the observer frame gives the $K$-correction for the respective spectroscopic epochs. 
Once the $K$ corrections are obtained from the sample at the spectroscopic epochs, they are also evaluated for the spectra of the SN for which 
the $K$-correction is to be determined. A temporal Gaussian Process (GP) regression 
is performed on the obtained $K$- correction values to make a prediction of $K$- correction for the desired photometric epochs of each SN with $z>0.01$. 
An example GP regression fit and the 68\% confidence level posterior prediction is shown in Fig. \ref{fig:Kcorrection} for SN~2004gy (top) and SN~2008bj (bottom).

\begin{figure*}
\begin{center}
\includegraphics[width=0.3\textwidth,height=0.2\textheight]{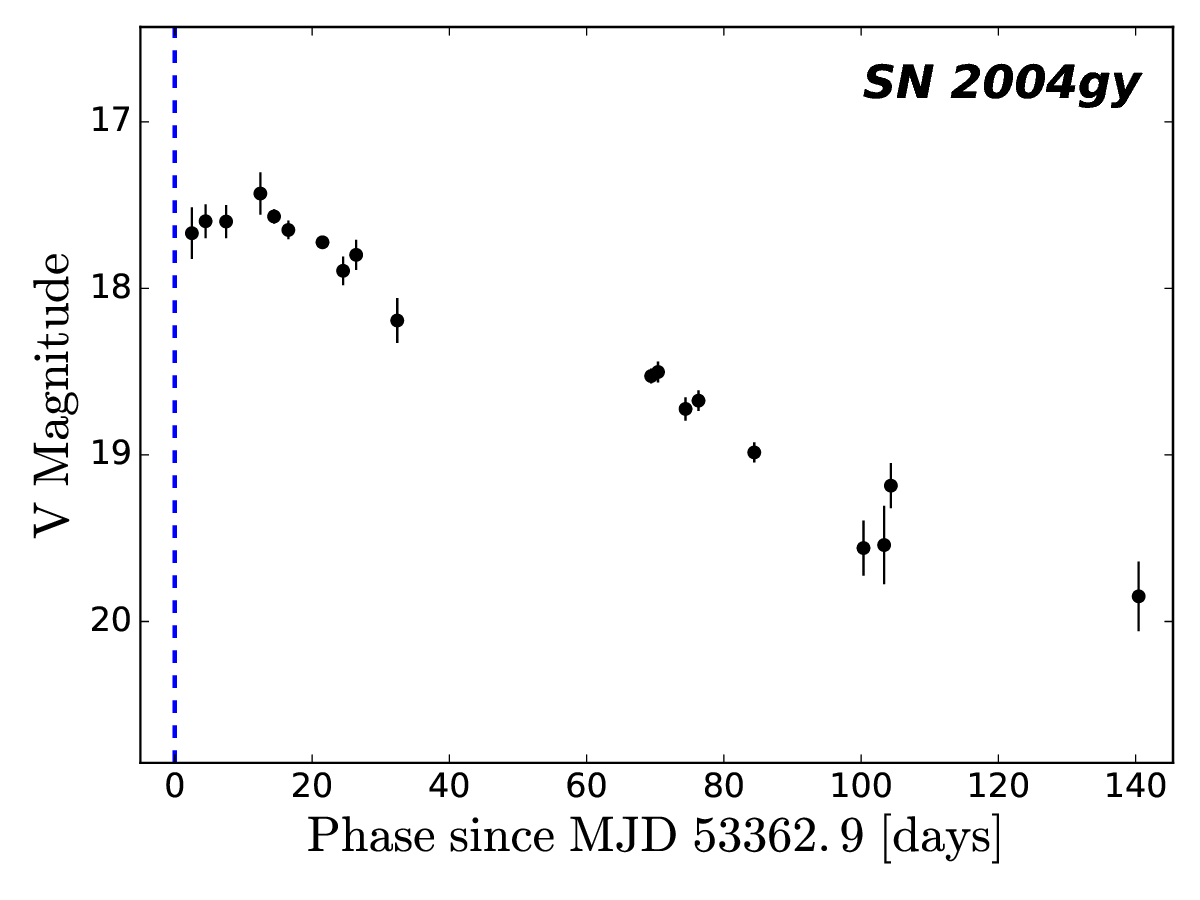}
\includegraphics[width=0.3\textwidth,height=0.2\textheight]{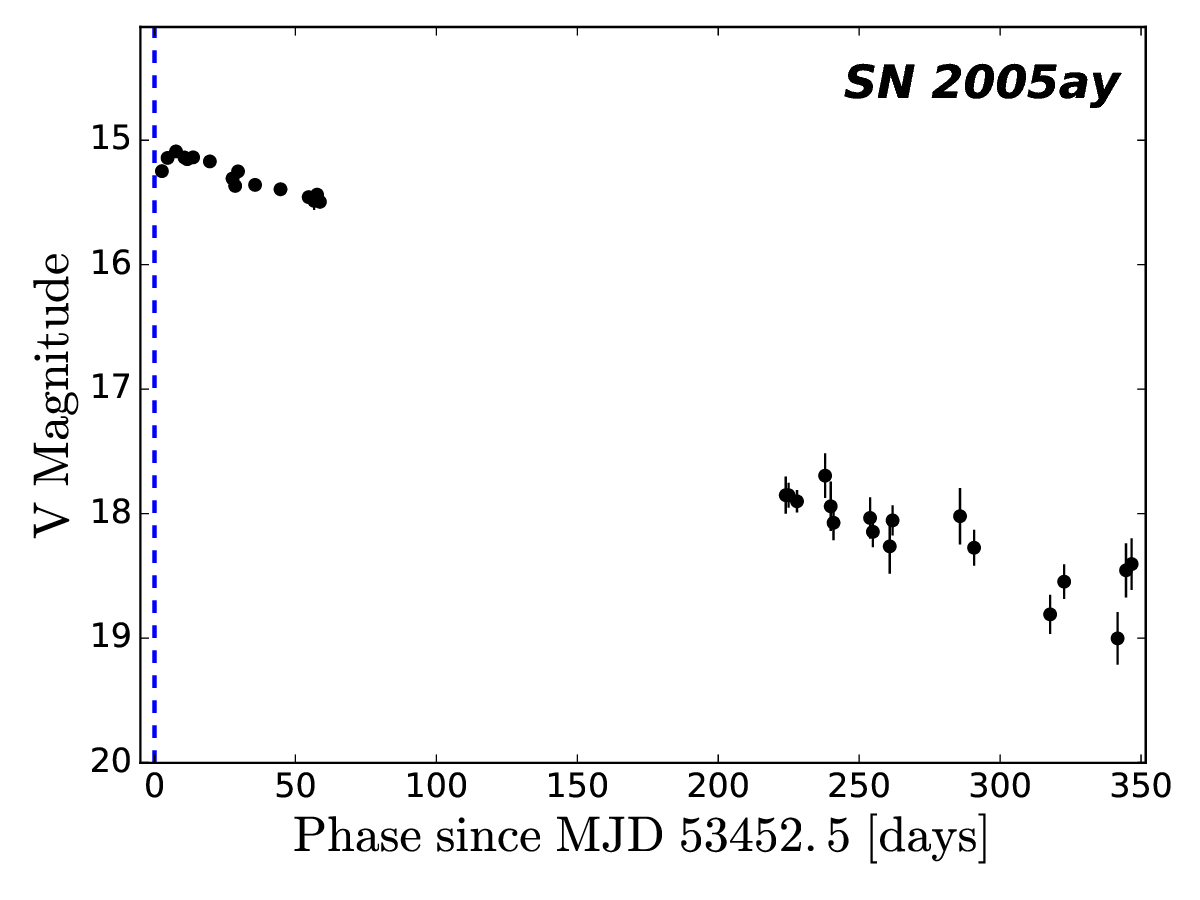}
\includegraphics[width=0.3\textwidth,height=0.2\textheight]{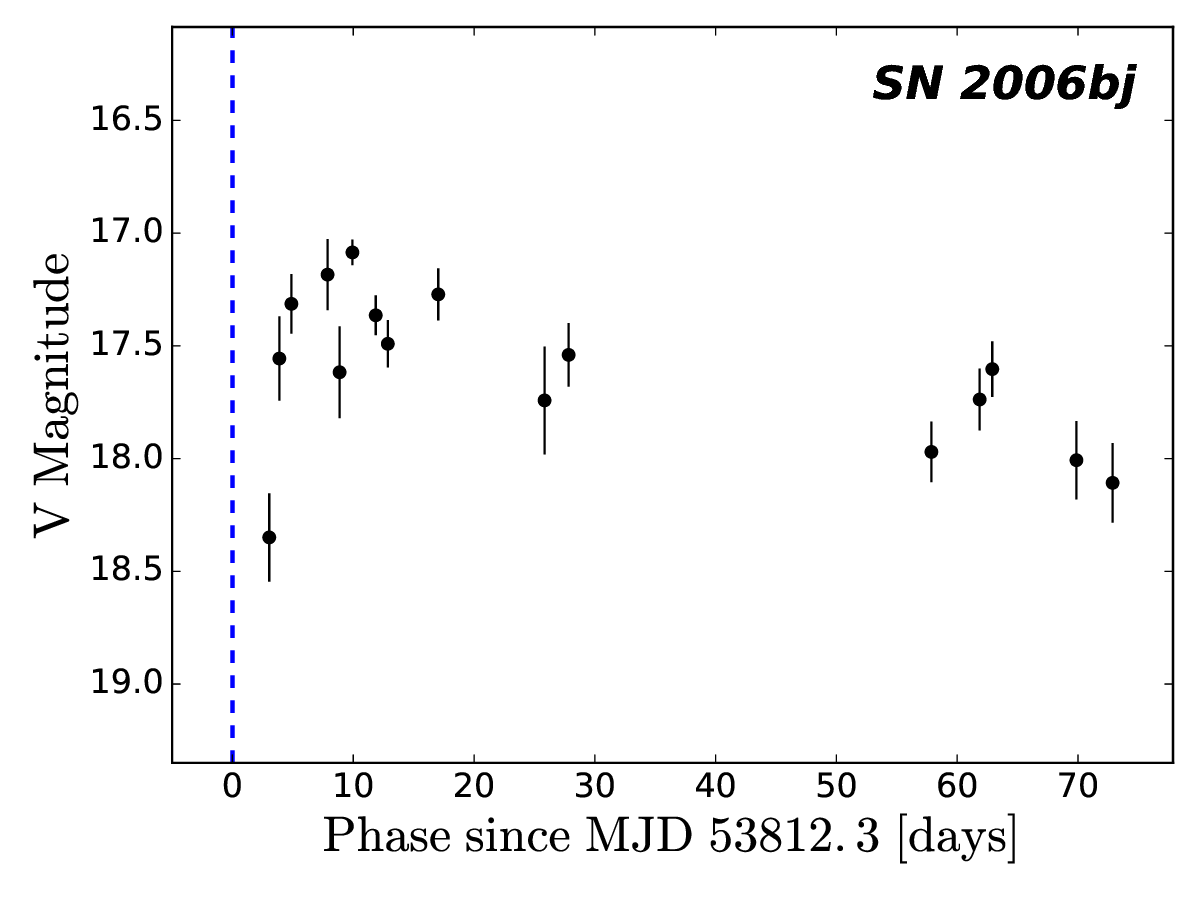}
\includegraphics[width=0.3\textwidth,height=0.2\textheight]{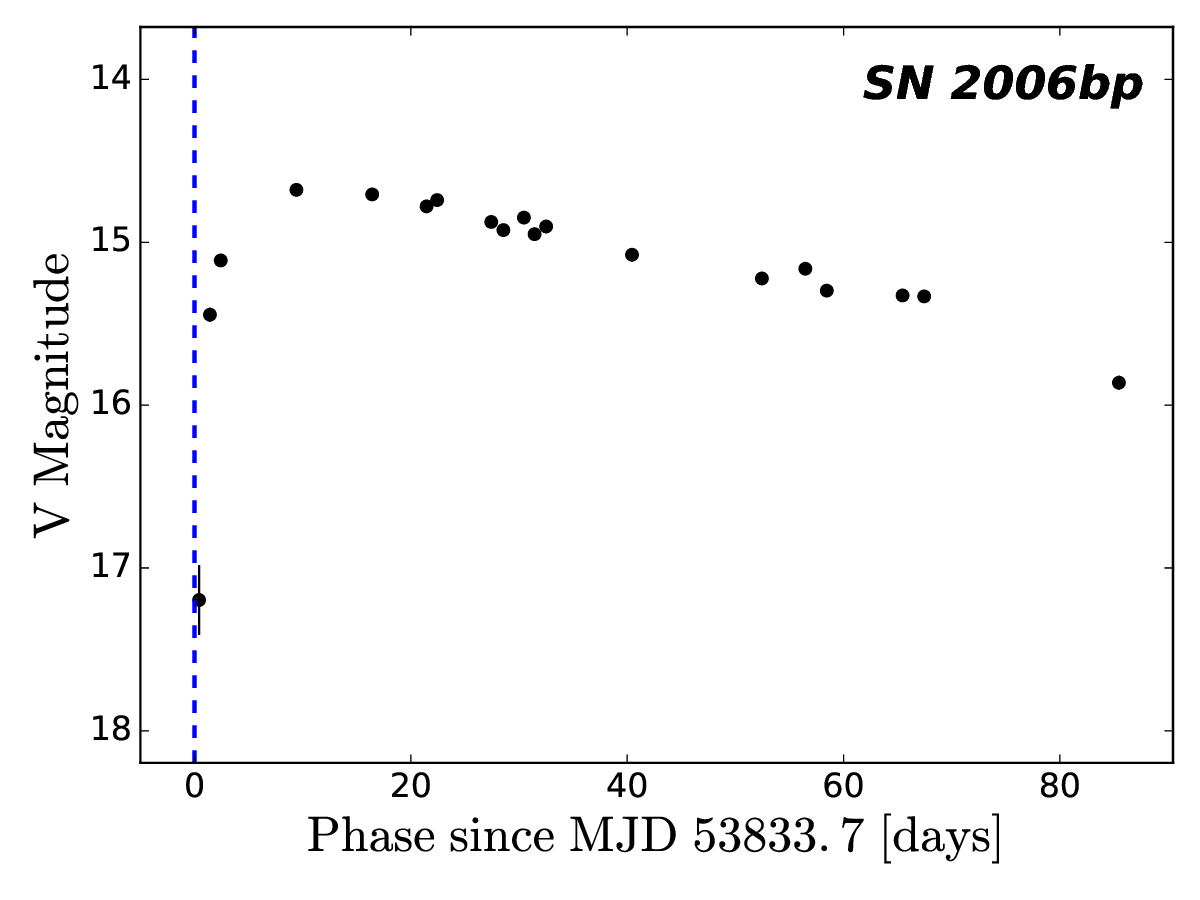}
\includegraphics[width=0.3\textwidth,height=0.2\textheight]{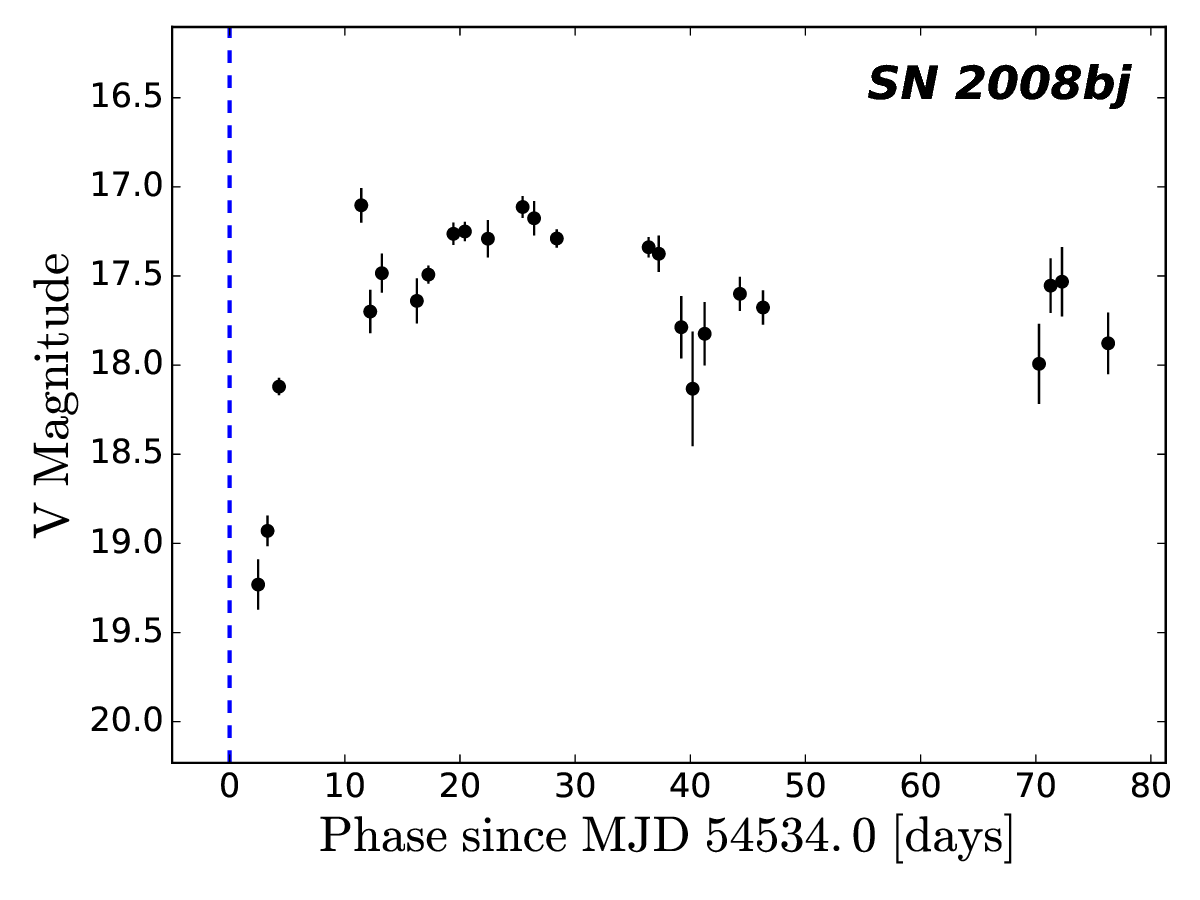}
\includegraphics[width=0.3\textwidth,height=0.2\textheight]{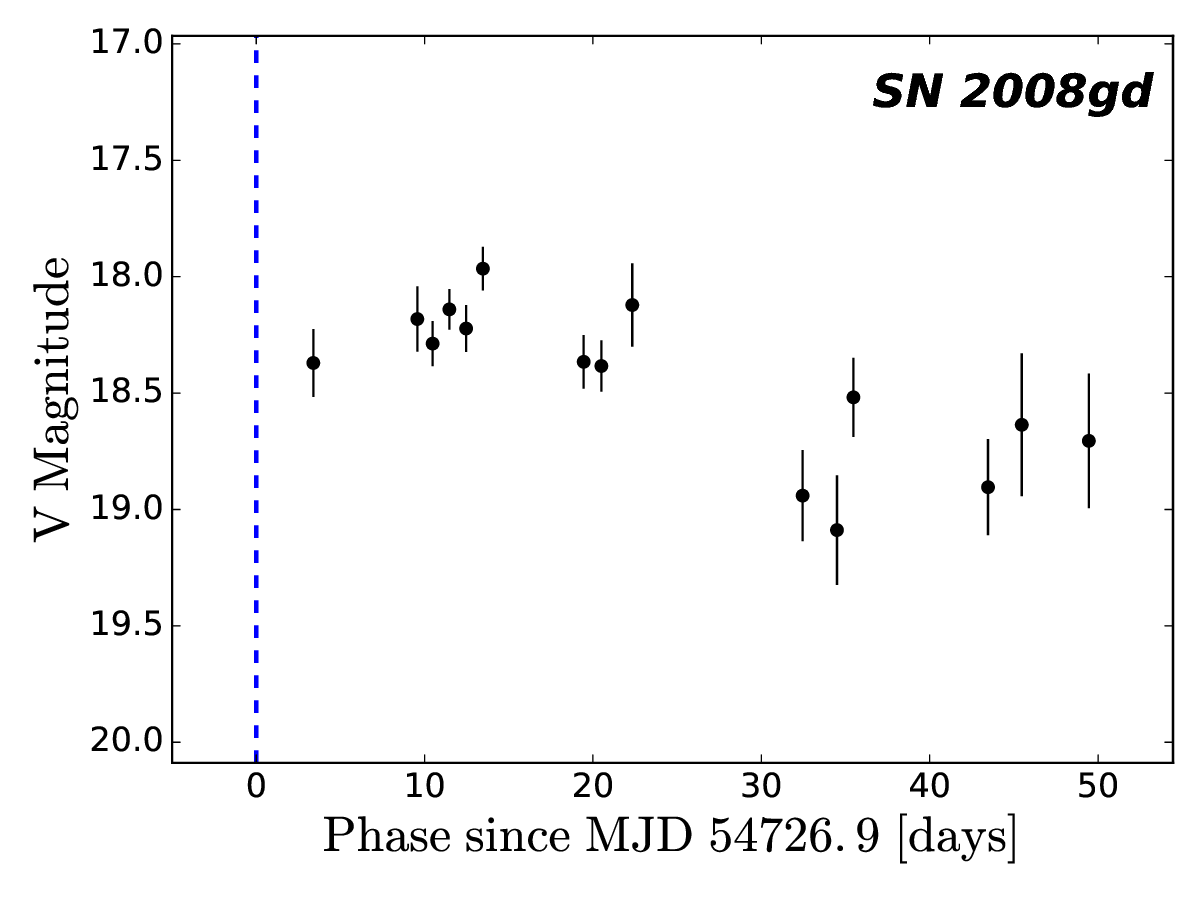}
\includegraphics[width=0.3\textwidth,height=0.2\textheight]{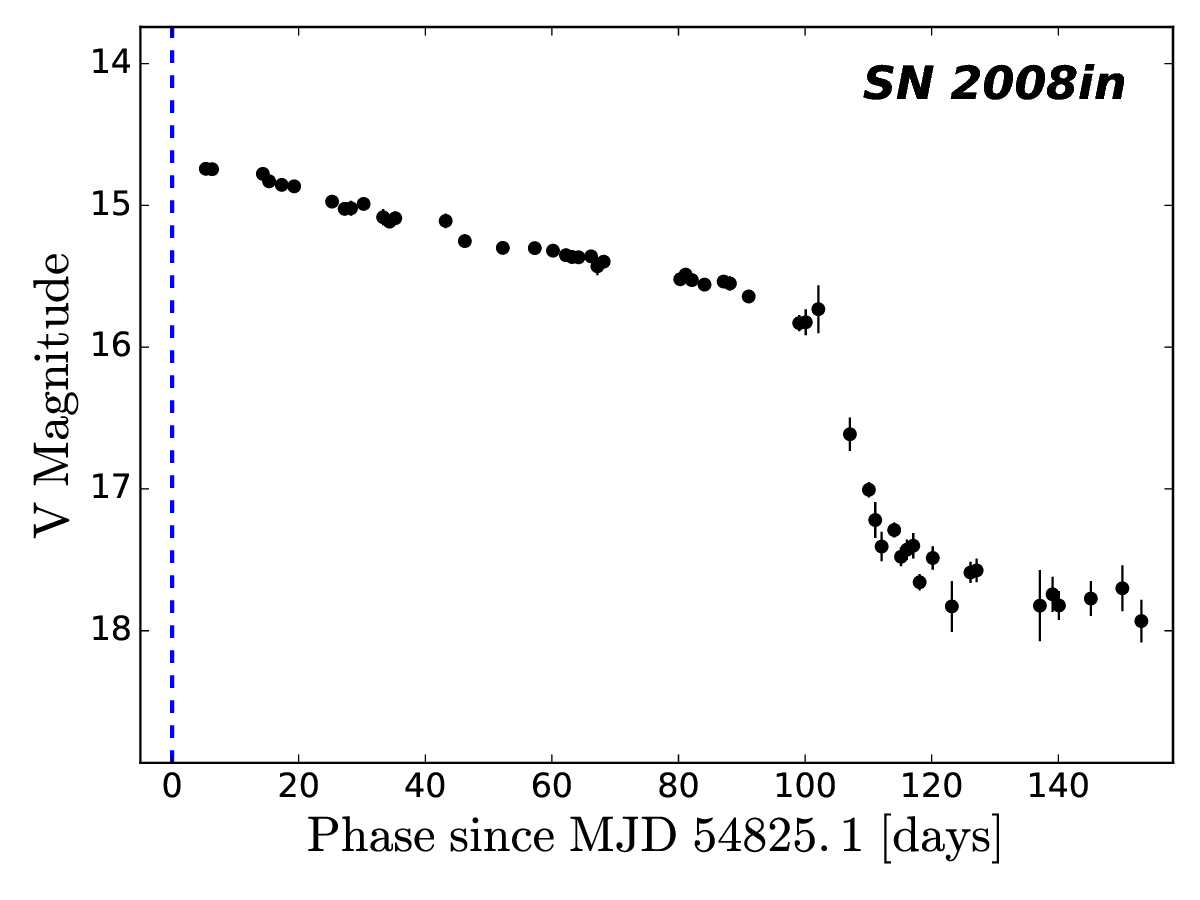}
\includegraphics[width=0.3\textwidth,height=0.2\textheight]{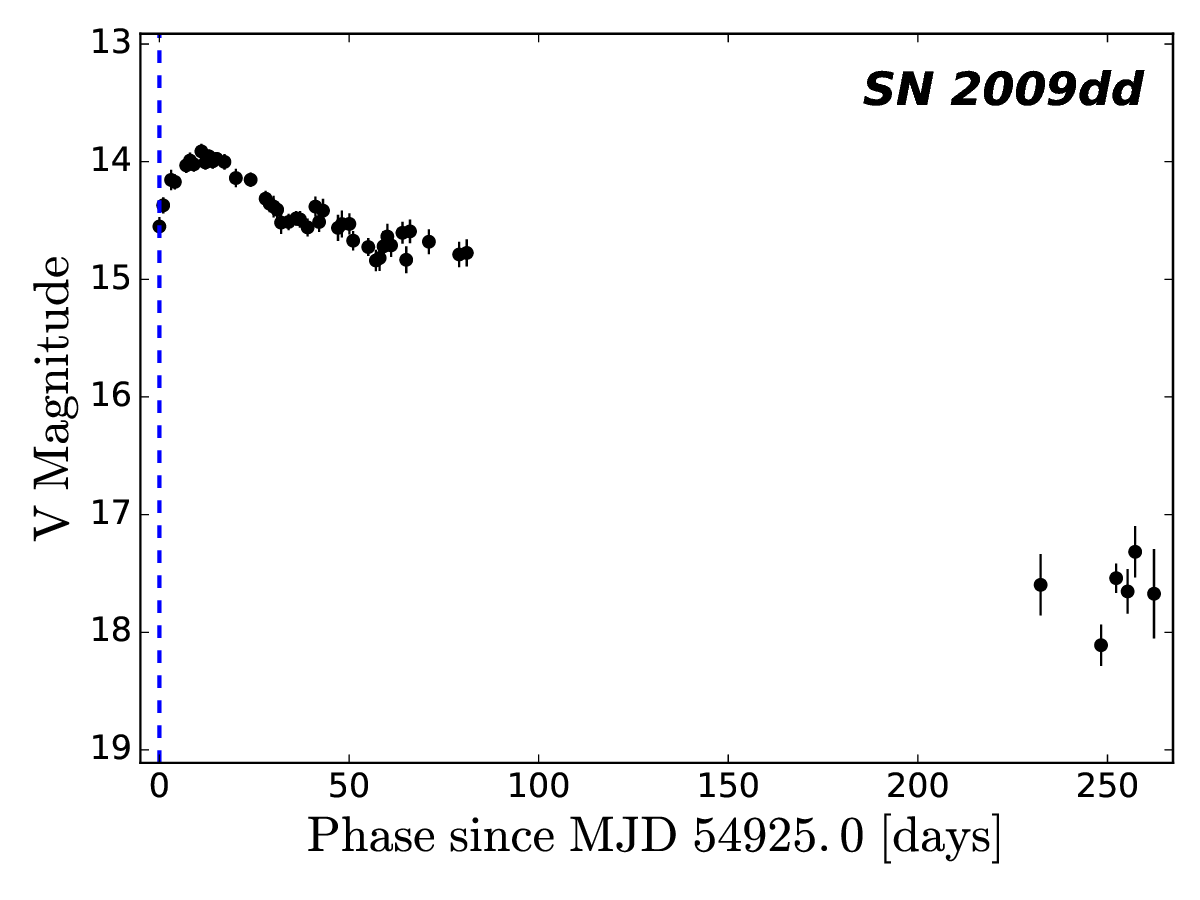}
\includegraphics[width=0.3\textwidth,height=0.2\textheight]{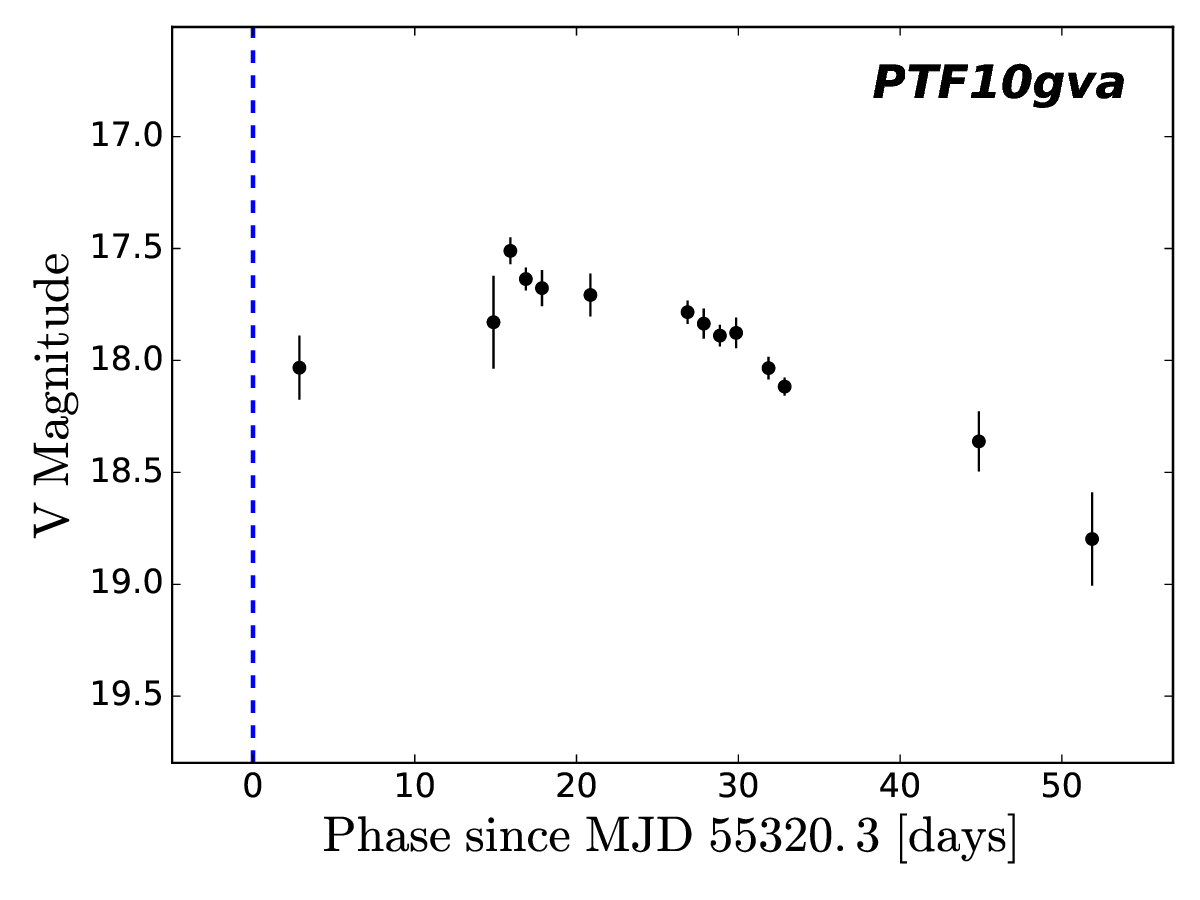}
\includegraphics[width=0.3\textwidth,height=0.2\textheight]{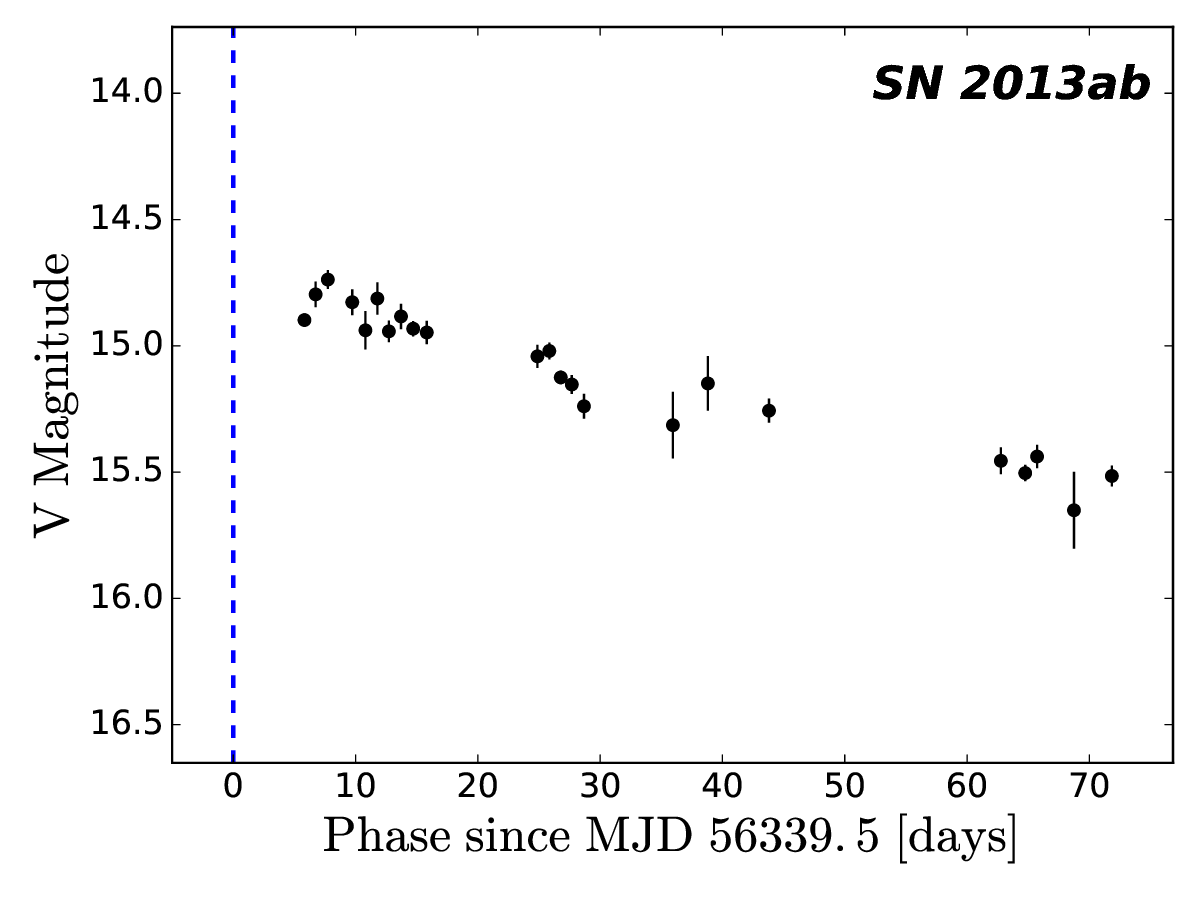}
\includegraphics[width=0.3\textwidth,height=0.2\textheight]{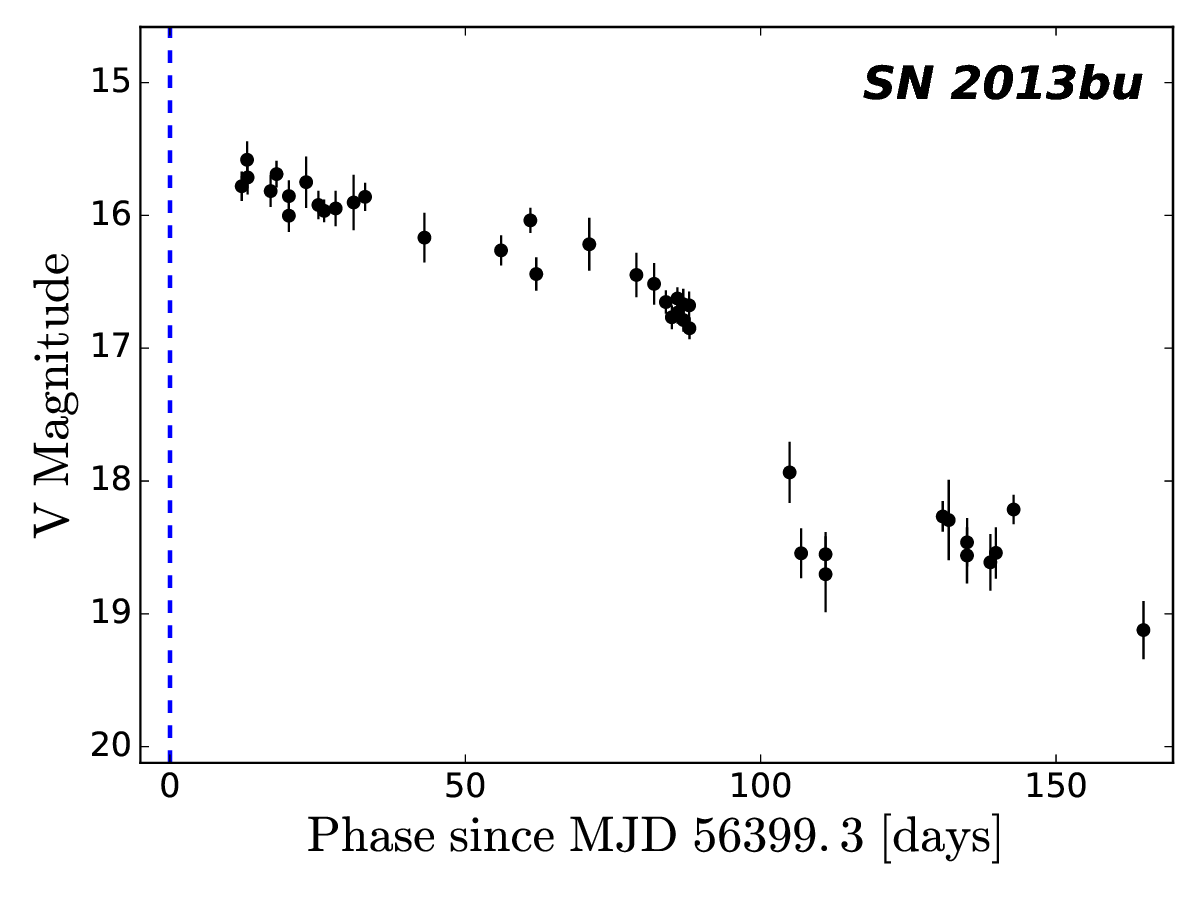}
\includegraphics[width=0.3\textwidth,height=0.2\textheight]{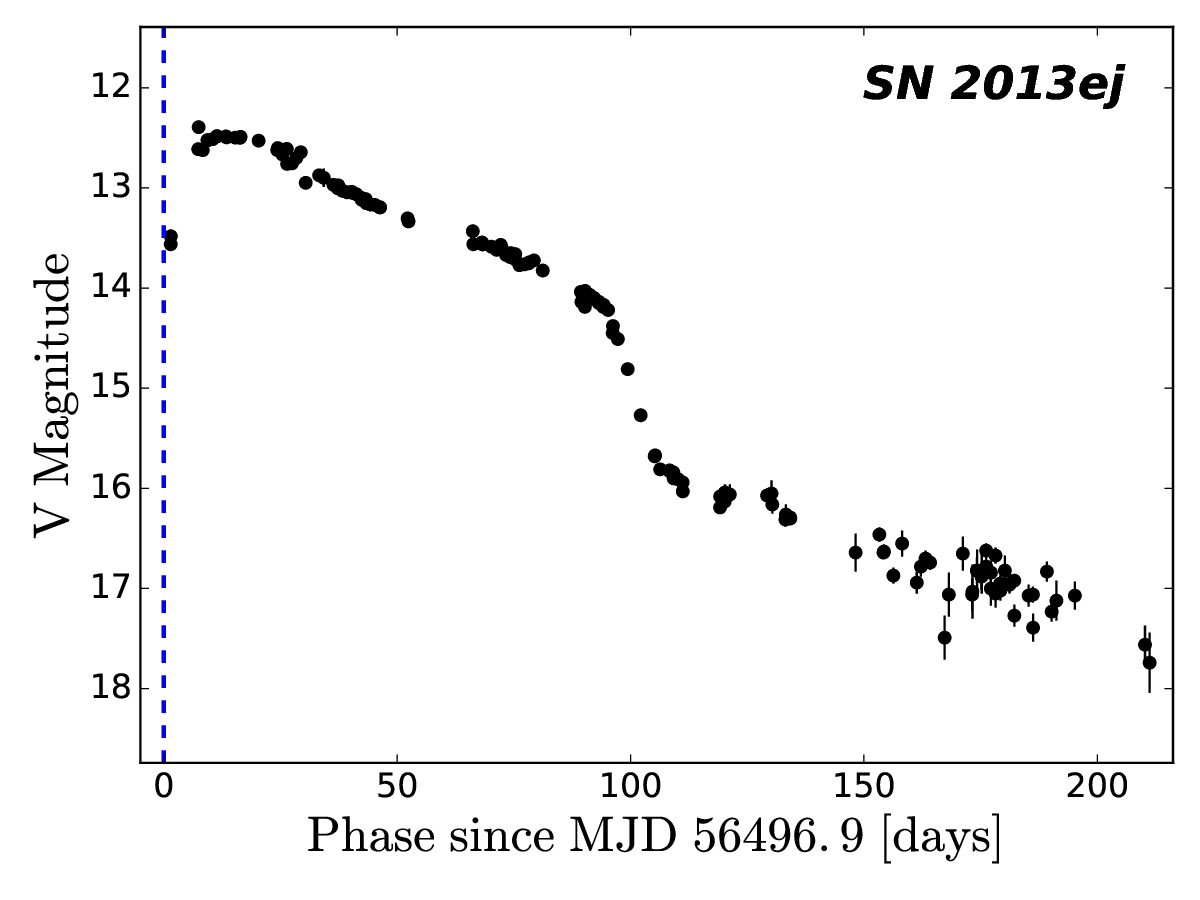}

\caption{Lightcurves of SNe in the ROTSE IIP sample after final photometric reduction and calibration. The ROTSE magnitudes are calibrated to APASS V band, followed by the SED correction using Eq.  \ref{eq:rotseVcalib} and the extinction correction. SNe with $z>0.01$ also include an additional $K$-correction. Blue dashed lines on each plot represents the adopted explosion epoch for the respective event.}
\label{fig:IIPsample_lc}
\end{center}
\end{figure*}

The final, fully reduced lightcurves for the ROTSE IIP SN sample, after calibration to APASS V magnitude and SED, and corrected for 
extinction and K-correction (for $z>0.01$) are shown in Fig. \ref{fig:IIPsample_lc}.

\section{Expanding Photosphere Method and Observables}\label{sec:epm}
With the advent of more sensitive, deeper surveys, SNe IIP discovery is increasing. While competitive samples to perform precise cosmological study 
at the higher redshifts are still accumulating, improvements in calibration for distance estimates have been explored using 
both photometry and spectroscopy in the lower redshift domain. We utilize the Expanding Photosphere Method (EPM) to estimate distances of our SNe IIP sample. We follow the prescription of \cite{dhungana16,DhunganaPhd} to describe the fundamental EPM equation, given by
\begin{equation}
t ~=~ {D \times \left ( {\theta \over v_{\rm phot}} \right ) + t_{0}},
\label{eq:epma}
\end{equation}
where $t$ represents the observing time, $D$ is the distance to the SN, $\theta = 2 R/D$ is the angular size of the photosphere at $t$, 
$v_{\rm phot}$ is the expansion velocity of the photosphere at $t$, and $t_0$ is the moment of the shock-breakout.

Assuming isotropic radiation from a blackbody, the observed flux can be written as
\begin{equation}
f_\lambda^{obs} = \theta^2\pi B_\lambda(T)10^{-0.4A_\lambda}
\label{eq:bbflux}
\end{equation}
where $B_\lambda(T)$ is the Planck function for the blackbody of effective temperature $T$. $A$ is the galactic extinction for the observed photometric band. The subscript $\lambda$ should be taken as an index for the observed photometric bands. Unlike a perfect blackbody, where the thermal photons emerge from the photosphere, the surface of last scattering (e.g., \cite{jones09, bose14}), SNe IIP photons are generated from the deeper atmosphere. Therefore, the parameter $\theta$ in Eq. \ref{eq:epma} corresponds to the thermalization layer while $v_{\rm phot}$ in Eq. \ref{eq:epma} corresponds to the photosphere  (optical depth, $\tau=2/3$) and the atmosphere is considered gray (e.g., \cite{jones09,eastman96}). A scaling factor $\zeta$, also termed the dilution factor or distance correction factor, is introduced as ratio of the radius of the thermalization layer to that of the photosphere.
\begin{equation}
\zeta = \frac{R_{\rm therm}}{R_{\rm phot}}
\end{equation}

Commonly, $\zeta$ is treated as a wavelength independent parameter in the optical and infra-red regime as described by \cite{eastman96}, 
who also show that it is a monotonic function of $T$ for several weeks after the explosion. 
Thus, to exploit EPM on SNe, care should be taken to select the measurement epochs when the wavelength dependence is not very significant. 
Complex computation of a realistic model atmosphere is required to accurately estimate $\zeta$, and is beyond the scope of this paper. 
We employ the commonly used prescription of \cite{dessart05}, and include $\zeta_\lambda$ in Eq. \ref{eq:bbflux}.
\begin{equation} 
\label{eq:epmtheta}
f_\lambda^{obs} =\zeta_\lambda (T)^2\theta^2\pi B_\lambda(T)10^{-0.4 A_\lambda}
\end{equation}

Ideally, one could consider full extinction-corrected bolometric flux by integrating over all the wavelengths. With the bolometric flux, $\theta$ can be obtained with
\begin{equation}  
\theta ~=~ {1 \over \zeta(T)} \sqrt{{f_{\rm bol}} \over {\sigma T_{\rm eff}^4} },
\label{eq:epmb}
\end{equation}
where $\sigma$ is the Stefan--Boltzmann constant. However, direct measurements of bolometric flux are not obtained in practice. Photometry is performed using specific pass bands. 
Thus, the filter response function is always convolved with the native flux from a SN, giving its magnitude in that wave band. Many SNe in the ROTSE IIP sample lack 
observations to yield or calibrate to the full bolometric flux. Therefore, we derive the effective blackbody flux by 
convolving with the filter response function $R_\lambda(\lambda)$. i.e.,
\begin{equation}
b_\lambda(T) = \int_0^\infty{R_\lambda(\lambda^{'})\pi B(\lambda^{'},T) d\lambda^{'}}
\end{equation}
Therefore, with the observed flux in a given pass band $\lambda$ and for the given value of $\zeta_\lambda(T)$, $\theta$ can be obtained using 
\begin{equation}
f_\lambda^{obs} = \zeta_\lambda (T)^2\theta^2 b_\lambda(T)10^{-0.4A_\lambda}
\label{eq:bbfluxfinal}
\end{equation}

The observed fluxes here should be treated as the $K$-corrected flux, whereas the parameters $\zeta_\lambda$ and $b_\lambda$ are in the SN rest frame. 
$K$-correction accounts for the (1+$z$) factors that would appear in the equations for high redshift SNe. Thus the derived distance will be the 
luminosity distance and not the angular diameter distance. We refer readers to \cite{gall16} and references therein for further discussion. 
The distance can be estimated using Eq. \ref{eq:epma} by determining $v_{\rm phot}$ and $T$, which can be directly obtained from observations. 
Then, both the parameters $D$ and $t_0$ can be simultaneously obtained by minimizing the $\chi^2$ using
\begin{equation}
\label{eq:epmchi}
\chi^2=\sum_j{\frac{ [\frac{\theta_{j}}{v_{{\rm phot}_{j}}}-\frac{(t_j - t_{0})}{D}]^2 }{{\sigma_j}^2}}
\end{equation}
where $\sigma_j$ is the uncertainty on $\theta_j/v_{{\rm phot}_{j}}$. 

\section{SNe IIP properties}\label{sec:prop}
The EPM distance estimation from Eq. \ref{eq:epmchi} now requires for each event 
a sample of $v_{\rm phot}$ and $\theta$ measurements. While $v_{\rm phot}$ at any epoch can be estimated from the 
line profiles in a spectrum, $\theta$ is obtained through temperature estimation and 
comparision with the concurrently observed photometric flux using Eq. \ref{eq:epmtheta}. 
In practice, however, both the spectroscopic and photometric measurements do not occur concurrently. 
Based on the available data, we will establish below the interpolation/extrapolation models to estimate the parameters at the desired epochs.
\subsection{Explosion Epoch}
Whenever available, the moment of explosion ($t_{0}$) for each of the IIP SNe sample is adopted from the literature as noted in Section \ref{sec:exporp}. 
When no estimate is available, we take $t_{0}$ to be the midpoint of the first of the ROTSE or a publicly available 
photometric detection, and the most recent pre-discovery non-detection epoch in the ROTSE data. We propagate the difference of $t_{0}$ and 
the pre-discovery epoch to the systematic uncertainty in $t_{0}$. Both the adopted explosion epoch 
and the respective uncertainty are given in Table \ref{tab:IIpsample}.

\subsection{Photospheric Velocity}
As the SN IIP ejecta exhibit an extensive H-envelope during early times, the photospheric velocities are generally estimated directly from the weak line 
signatures such as those of Fe II lines. While in the plateau phase, the absorption 
minima of the P Cygni profiles of Fe~II $\lambda4924$, $\lambda5018$ and $\lambda5169$ have been 
used as the best estimators of the photospheric velocity 
(e.g \cite{leonard02, dessart05, nugent06}), in the earlier epochs these 
lines cannot be observed. 
When observed, we strictly use Fe~II $\lambda5169$ velocity measurements for the $v_{\rm phot}$. When Fe~II $\lambda5169$ is not observerd at early times, we use He II $\lambda5876$ line. On rare occasions, when He II also are not observed, we use correlations of H Balmer line velocity with Fe II line velocity from \cite{Faran14} to obtain an effective $v_{\rm phot}$.
\subsubsection{Measurement}
\begin{figure*}
\begin{center}
\includegraphics[width=\textwidth]{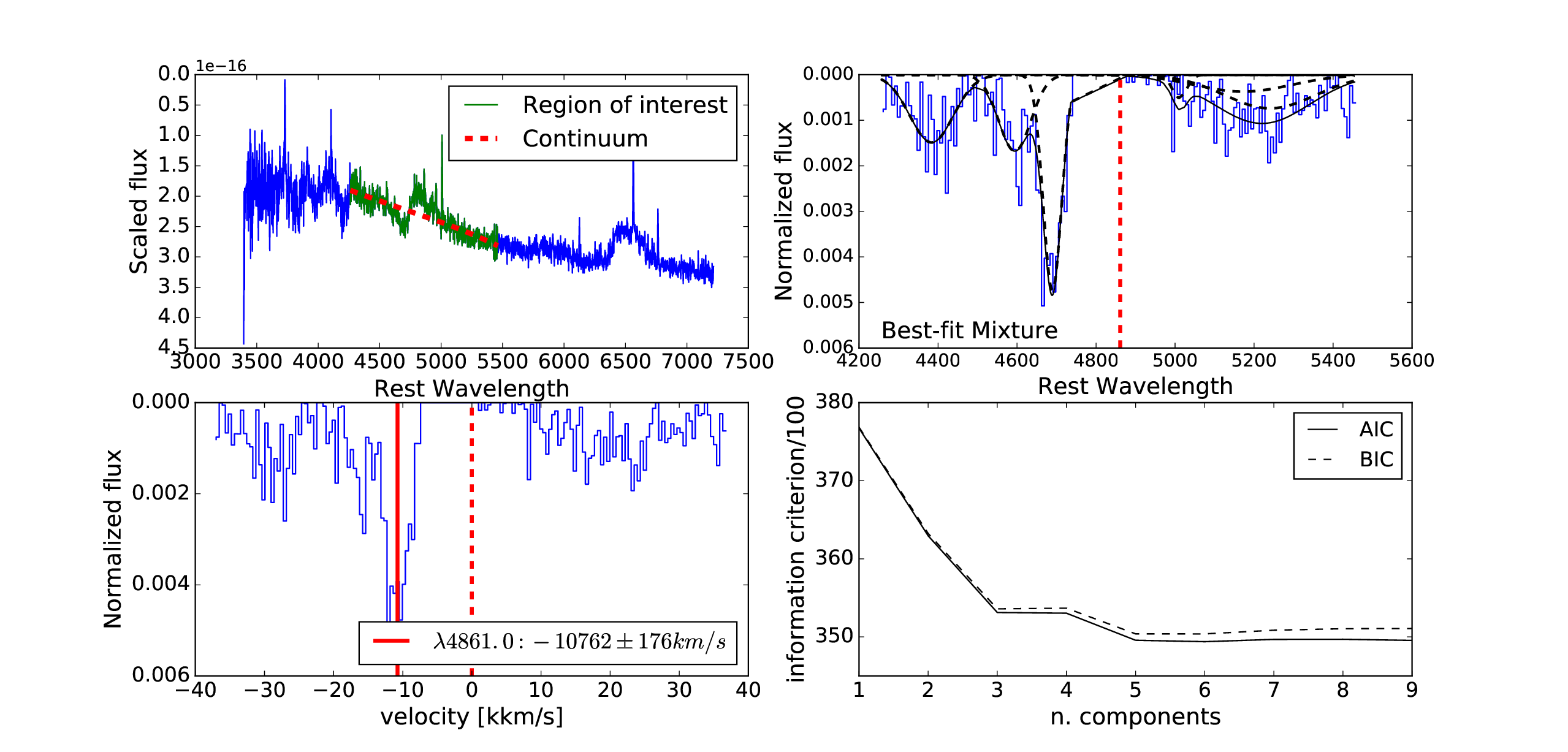}
\caption{Estimation of the velocity of the H$\beta$ line using the GMM model. Top Left: A spectrum of SN 2004gy taken on Jan 10, 2005. The continuum is 
estimated using a spline fit on the side bins around the signal region of 4500 -- 5000~\AA. Top Right: The normalized residuals on the continuum 
subtracted absorption profile. The best fit mixture model is shown in solid black and the respective 6 Gaussian components are shown in dashed black lines. 
The dashed vertical line is the rest frame line position for H$\beta$. Bottom Left: Estimation of the line velocity from the best fit. The dashed 
vertical line indicates the rest frame position of H$\beta$ and the solid line shows the position of the absorption 
minimum obtained from the best fit. Bottom Right: BIC and AIC estimation for mixture model with multiple number of Gaussian components ($n$) 
performed iteratively for $n$= 0 -- 9. The best fit has minimum BIC and corresponds to n=6}
\label{fig:HBetavel}
\end{center}
\end{figure*}

A convenient way to estimate the position of the line minimum 
is to perform a Gaussian fit of the absorption profile. Accurate measurements are complicated by typically blended features and continuum subtraction. 
We perform a 1D Gaussian mixture modeling (GMM) of a segment of the spectrum 
around each line of interest. We define a signal region considering the whole line profile.  
We consider a few 100s of Angstroms on both sides of the signal region as the side bins and a 
continuum is estimated by performing a spline fit on the side bins. 
GMM is performed iteratively on the continuum subtracted absorption profile. 
The best fit GMM model gives the optimum number of the Gaussian components as the maximum likelihood fit of this absorption profile. We measure 
the Bayesian Information Criterion (BIC) and Akaike Information Criterion (AIC) for each model 
as we increase the number of components for each iteration. The model that yields the minimum BIC 
from the GMM fit is chosen to be the best model. We monitor AIC also to ensure that it does not severely contradict with the best model from the minimum BIC value. The position of the component aligned with the 
line of interest will give the best fit value and the uncertainty of the photospheric velocity. 
An example involving this process of estimating line velocity for the H$\beta$ line taking a 
spectrum for {SN 2004gy} from Jan. 10, 2005 is shown in Fig. \ref{fig:HBetavel}. The top left panel shows the estimation of the continuum obtained 
from the spline fit performed on the side bins around the signal region of 4500 - 5000~\AA. On the top right is shown the normalized 
residuals on the absorption profile after subtracting the continuum. GMM is performed iteratively by varying the number of components from 0--9. 
The bottom right panel shows the corresponding BIC and AIC values for each mixture model. The best fit model with the minimum BIC has six Gaussian components 
and the corresponding model and the Gaussian components are shown on the top right panel. The line velocity is obtained from the minimum of 
the Gaussian component corresponding to H$\beta$ and is estimated to be $10760\pm 176~{\rm km/s}$, shown on the bottom left panel.
%While Fe II lines are not observed during the early times, the estimated Balmer lines and He II 5876 line 
%velocities are transformed to effective photospheric velocities using the correlations of \cite{Faran14} (see Fig. 16 in their paper).

\subsubsection{Extrapolation Model} \label{sec:vphotmodel}
Once the ionic velocities are measured directly from the GMM on the observed spectra, 
we want to interpolate/extrapolate these measurements to the photometric observation epochs. 
Previous studies such as \cite{poznanski10,Faran14} have empirically modeled the temporal evolution of 
prominent ionic signatures such as from Fe II lines. While Fe II lines tend to track the photospheric velocity, 
these lines are generally unobserved during the first few weeks. We show below that extrapolating 
such a model to early times shows a steeper 
decline than the velocities directly observed from He II lines. 

\begin{figure*}
\begin{center}
\includegraphics[width=\textwidth]{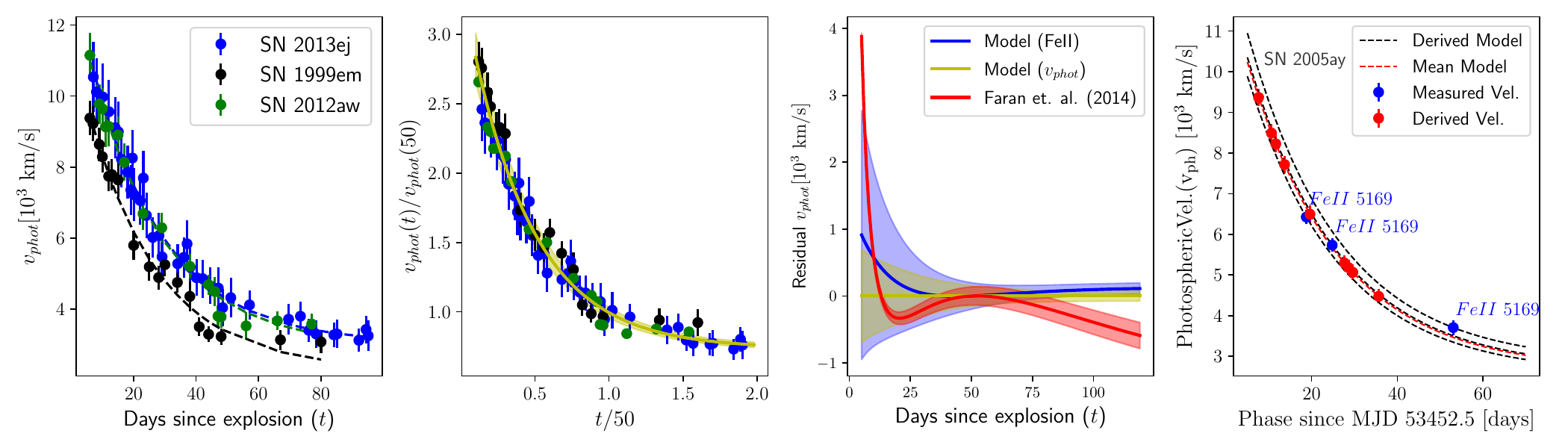}
\caption{Left: Measured photospheric velocities and the evolution for three well studied SNe IIP. 
Each velocity is derived from fitting relevant spectral lines using the GMM model from the spectra at corresponding epochs. 
The color-coded dashed lines represent the exponential decay law fits using Eq. \ref{eq:vphotIIp} 
for the respective events. Middle Left: The global decay law fit after calibrating the epochs and $v_{\rm phot}$
relative to 50 day values for each of the SNe. The best fit model, shown with the solid line, has a $\chi^2/dof = 0.60$.
Middle Right: The residual velocity evolution models after subtracting the best fit $v_{\rm phot}$ model shown in the middle left panel. The residual 
obtained for model obtained using Fe II lines alone is shown in blue and the red line is the residual obtained for the Fe II model evolution from \cite{Faran14}. All three are shown
relative to a arbitrary pivot point of  $v_{\rm phot}(50)=4000$ km/s. Right: Measured and derived $v_{\rm phot}$ models for SN~2005ay using the best fit model. The blue points 
are the labelled ion velocities measured directly from the spectra using GMM. The black dashed curves 
are the respective $v_{\rm phot}$ models calibrated to the three blue points. The red dashed curve is the model calibrated to the weighted mean velocity $v_{\rm phot}(50)$ and red points are the derived estimates at the SN photometric epochs.}
\label{fig:vphotevolve}
\end{center}
\end{figure*}

We assemble the line 
velocities of three well sampled supernovae from the literature and observe the time series evolution of the $v_{\rm phot}$. 
We first test an exponentially decaying model that appears to closely capture the $v_{\rm phot}$ evolution for each 
event during the epochs considered. Furthermore, when the epochs and $v_{\rm phot}$ of each SNe are 
calibrated relative to 50 day values, and the exponential model is fitted on the full distribution, 
we observe the RMS scatter to decrease significantly, yielding a reasonable $\chi^2/dof$ of 0.60 for the fit. We note that 
the choice of 50d has been commonly made in the literature and for SNe IIP, this is about midpoint of the 
typical plateau phase where the evolution is relatively smooth. We use an exponentially decaying model for the 
velocity evolution, given by
\begin{equation}
\label{eq:vphotIIp}
%\frac{v_{\rm phot}(t)}{v_{\rm phot}(50)}= a+b~e^{c(\frac{t}{50})}
\frac{v_{\rm phot}(t)}{v_{\rm phot}(50)}= a+b~\exp[-c \cdot (t / 50) ].
\end{equation}
The best fit parameters for the velocity evolution in Eq. \ref{eq:vphotIIp} are estimated to be 
$a=0.735\pm0.025$, $b= 2.650\pm0.064$; and $c=2.327\pm0.589$. The left panel of Fig. \ref{fig:vphotevolve} 
shows the time evolution of the $v_{\rm phot}$ for three SNe obtained from the GMM fit of the  absorption 
profile of ions during the photospheric period.
The middle left plot shows our best fit model after calibrating relative to 50 days. The middle right plot shows the 
residuals for different models after subtracting $v_{\rm phot}$ evolution best fit model shown in the middle left panel. Blue line is the residual for velocity model, obtained using Fe II only lines extrapolated in the earlier times, after subtracting best fit $v_{\rm phot}$ model. Also shown in red correspond to the residual for Fe II evolution model from \cite{Faran14} extrapolated at early times, after subtracting the same best fit model. 
For comparison, all three models are anchored at $v_{\rm phot}(50)=4000$ km/s and the respective 
shaded region indicates the 68\% confidence region. We see the Fe II only model 
yields a steeper velocity evolution than the $v_{\rm phot}$ model using He II lines at early times. 

To estimate the $v_{\rm phot}$ for our IIP sample, we first measure line velocities directly from the obtained spectra of the respective SNe. 
We apply the $v_{\rm phot}$ evolution model Eq. \ref{eq:vphotIIp} to obtain the $v_{\rm phot}(50)$, and use it as anchor to sample the 
velocities at the desired photometric epochs.
For events with multiple spectra, a weighted mean of $v_{\rm phot}(50)$ is obtained 
and the $v_{\rm phot}$ at the photometric epochs are obtained anchoring the model Eq. \ref{eq:vphotIIp} using this average $v_{\rm phot}(50)$. 
We propagate the statistical error as uncorrelated systematic uncertainty from $v_{\rm phot}$ into the distance error budget for each SN. 
An example showing the directly estimated $v_{\rm phot}$ and the evolution model 
for the SN~2005ay is shown in the right most plot of Fig. \ref{fig:vphotevolve}. The blue points are the measured 
velocities for the labelled ions. The black dashed curves are the respective model pivoted at the measurement epoch, while 
the red dashed curve is the model anchored at the average $v_{\rm phot}(50)$ measurement and red points are the velocity estimates at the photometric epochs of the SN.  
We perform the same procedure for all 12 events in our SNe IIP sample. 
We propagate the uncertainty on $v_{\rm phot}$ due to the evolution model as a 100\% correlated uncertainty across all 12 SNe in the sample.

\subsection{Temperature} \label{sec:temp}
The parameter $\theta$ in Eq. \ref{eq:epmchi} is obtained by comparing the observed 
flux from the SN with the effective blackbody flux after accounting for the 
dilution correction as shown in Eq. \ref{eq:epmtheta}. To determine the effective 
blackbody flux, we would like to estimate the temperature from fitting the SED constructed from 
the $BVI$ phototmetric measurements to a Planck function.
\subsubsection{Measurement}
For the events where $BVI$ observations are available, we fit the measured SED to 
a Planck function directly and temperature is directly obtained as a fit parameter.
For other events, to estimate the effective color or blackbody temperature, we take 
the spectra and spectro-photometrically determine the $BVI$ fluxes. For this, each redshift 
corrected spectrum is unreddened for the galactic extinction by applying the reddening curve 
using the parametrization from \cite{Fitzpatrick99}. The color extinction $E(B-V)$ is obtained 
for respective SNe from the literature, while \cite{sf11} is used for the Milky-way extinction if no such information is available. After this, we derive a set of synthetic $BVI$ magnitudes 
at the rest frame, and the resulting SED is fitted to a Planck function to estimate the effective 
temperature as before. For cases where the spectral coverage is not wide enough to synthesize 
the $BVI$ fluxes, the corrected spectrum is fitted directly to the Planck function after 
masking the H$\alpha$ line and the telluric lines from the atmosphere when present. 

\subsubsection{Extrapolation Model}
\begin{figure*}
\begin{center}
\includegraphics[width=\textwidth]{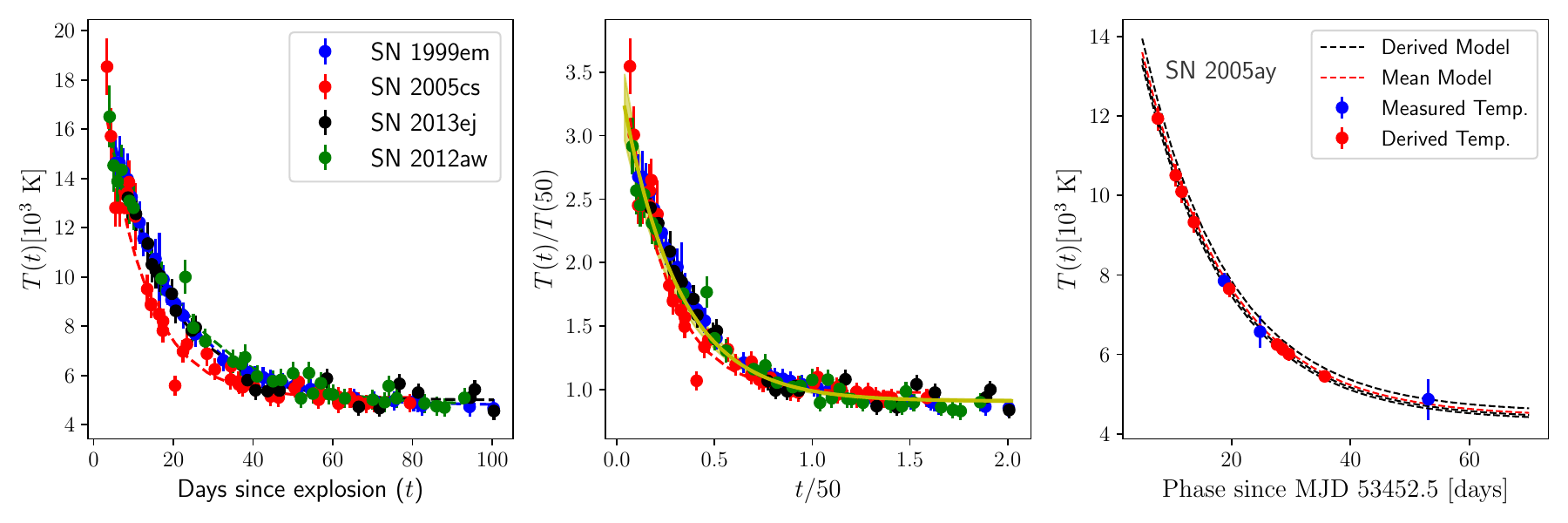}
\caption{Left: Measured color temperature and the evolution model for four well studied SNe IIP. 
Each temperature is derived from fitting $BVI$ fluxes to Planck function. The color-coded dashed lines 
represent the exponential decay law fits using Eq. \ref{eq:tempIIp} for the respective events. 
Middle: The global decay law fit after calibrating the epochs and temperature relative to 50 day values 
for each of the SNe. The best fit model, shown with the solid line, has a $\chi^2/dof = 1.04$. Right: Evolution of effective temperature for SN~2005ay. Blue points are the measured temperatures 
obtained from SED fit to Planck function.
The dashed black lines are the model extrapolation anchored at each blue point, while the red dashed line 
is the effective model pivoted at the weighted average T(50) value. 
Red points are the sampled model temperatures at the photometric epochs.}
\label{fig:tempevolve}
\end{center}
\end{figure*}

To estimate the temperature at the ROTSE photometric epochs, we first establish an analytic 
relation of the temperature evolution empirically using a sample of very well observed SNe IIP. 
As shown in the left panel of Fig. \ref{fig:tempevolve}, an exponentially decaying model appears to 
accurately capture the temperature evolution for each event during the epochs considered. 
We pursue an approach to generate our temperature evolution model analogously to the velocity 
evolution model described in Section \ref{sec:vphotmodel}.
We translate the epochs and temperatures of each SNe relative to 50 day values and perform an 
exponentially decaying model fit on all three SNe. We observe the RMS scatter to decrease 
significantly as shown in the middle panel of Fig. \ref{fig:tempevolve}. 
The choice of 50 day epoch is 
consistent with the velocity evolution model. The exponential model for the temperature evolution is given by
\begin{equation}
\label{eq:tempIIp}
%\frac{T(t)}{T(50)}= a+b~e^{c(\frac{t}{50})}
\frac{T(t)}{T(50)}= a+b~\exp[-c \cdot (t / 50) ]
\end{equation}
The best fit model for the full sample yields $a=0.908\pm0.012$, $b= 2.662\pm0.091$ and $c=3.492\pm0.143$, as shown in Fig. \ref{fig:tempevolve}, and yields a $\chi^2/dof=1.04$.  We also note that a power law model with decay index of $-0.44\pm0.01$ also yields a reasonable estimate with a $\chi^2/dof=1.6$.  The adopted exponential model serves to reproduce the full temperature evolution if one temperature is 
measured at any epoch. The temperature estimated directly from the $BVI$ fluxes is used to estimate $T(50)$ using Eq. \ref{eq:tempIIp}. Similar to the photospheric velocity evolution, for events with multiple spectra, the final temperature evolution is derived from the weighted mean $T(50)$ value.  The estimated $T(50)$ is then used to anchor the model to desired photometric epochs. The evolving temperatures are sampled from this model at the ROTSE photometric epochs between +7d and +35d. The statistical uncertainty for each SN is propagated as an uncorrelated systematic uncertainty from the temperature evolution into the distance error budget.
The measured temperatures, the extrapolation models and the respective final estimate 
at the ROTSE epochs for SN 2005ay are shown in the right panel of Fig. \ref{fig:tempevolve}.  For each SN in 
the sample, we propagate the extrapolation model uncertainty as a correlated systematic uncertainty from this common temperature evolution model in the final distance error estimates. 
Table \ref{tab:epmmeasure2} shows the derived +50 day values of the photospheric velocity and the 
effective temperature for all 12 SNe in our IIP sample. 

\section{Distance Measurement}\label{sec:dist}
We have established the overall methodology and the various inputs that will go into our measurement of SN distance. The photometric measurements and their relationships with the EPM parameters are discussed in Section \ref{sec:epm}. The best fit distance for each SN is estimated by $\chi^2$ minimization using Eq. \ref{eq:epmchi}. We next identify systematic effects that could impact our results and assess their magnitude on the final measurements. As we have modeled both the temperature and the photospheric velocity anchored at the 50 day estimate, in each case, there are two categories of systematic uncertainties that are propagated to the final distance estimation. One is the uncorrelated systematic uncertainty from the uncertainty on the respective 50 day value, while the other is the correlated systematic uncertainty from the modeling of those parameters at the photometric epochs of each SNe. Both of these uncertainties are propagated to the distance systematic uncertainty.  The uncertainties in the galactic extinction (based on \cite{sf11}) and the host 
extinction $E(B-V)$ are also transformed as uncorrelated systematic uncertainty in distance. 
The additional systematic uncertainty from the $K$-correction modeling is also propagated from 
the posterior estimate from the Gaussian Process regression.  Finally, we also propagate the uncorrelated systematic uncertainties on the adopted $t_{0}$, which are given in Table \ref{tab:IIpsample}. Better constrained 
shock breakout times can significantly reduce the total uncertainty on the distance estimated with the EPM method. 
The fitted EPM distance and $t_{0}$ with the respective uncertainties from the fit are given in Table \ref{tab:epmmeasure2}. All epochs including the fitted values of $t_{0}$ are relative to the adopted $t_{0}$ values in MJD from Table \ref{tab:IIpsample}. We also note that the uncertainties on $t_{0}$ from the fit are consistent with the our adopted uncertainties in Table \ref{tab:IIpsample}.

\begin{figure*}
\begin{center}
\includegraphics[width=0.3\textwidth,height=0.2\textheight]{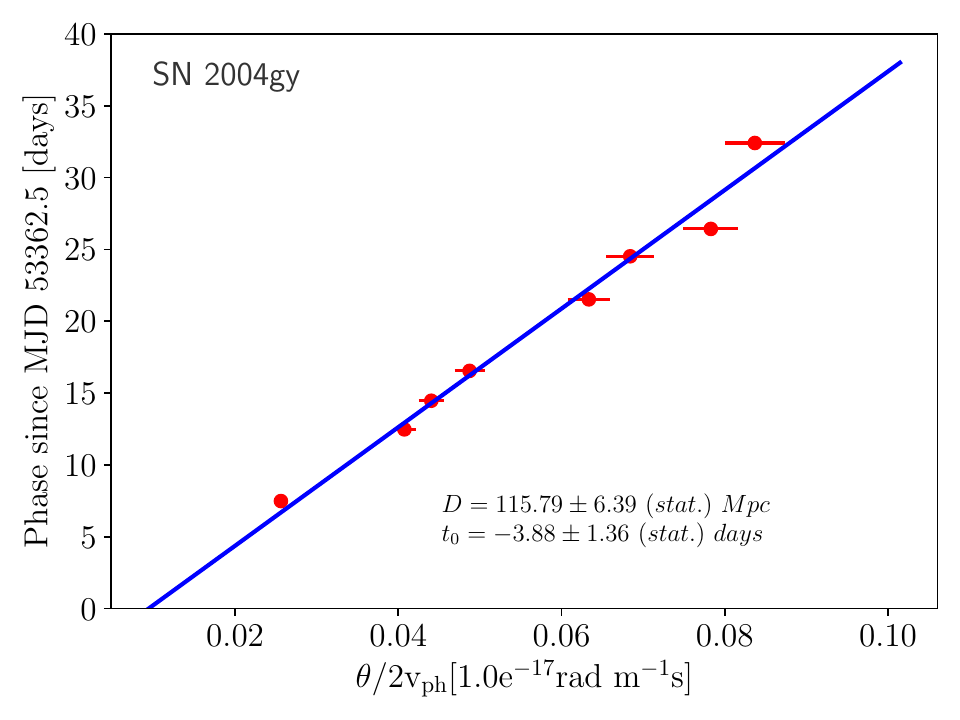}
\includegraphics[width=0.3\textwidth,height=0.2\textheight]{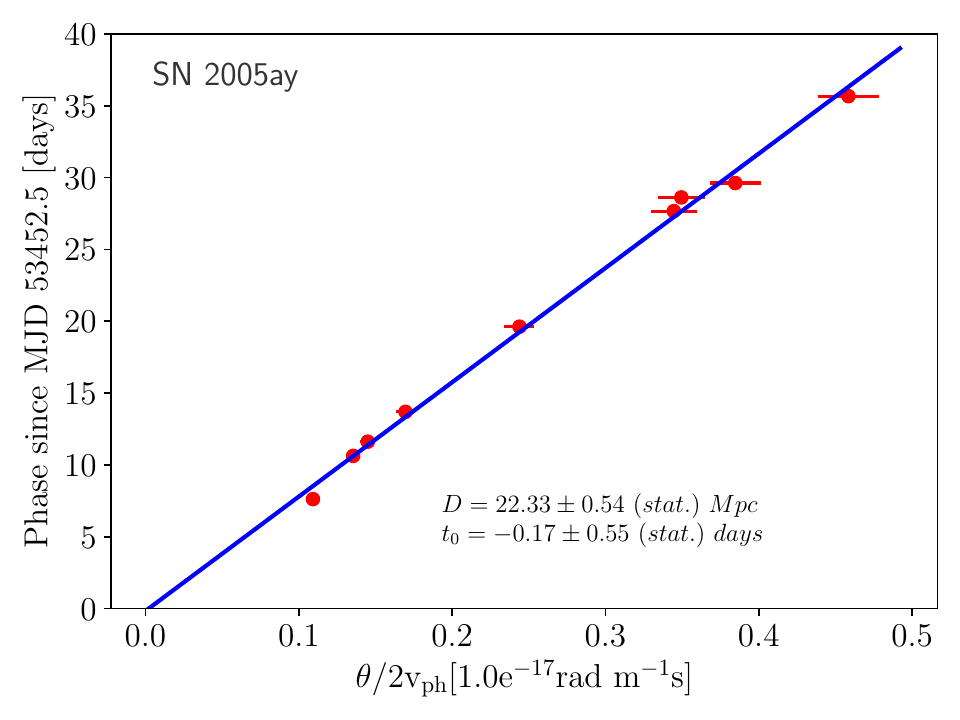}
\includegraphics[width=0.3\textwidth,height=0.2\textheight]{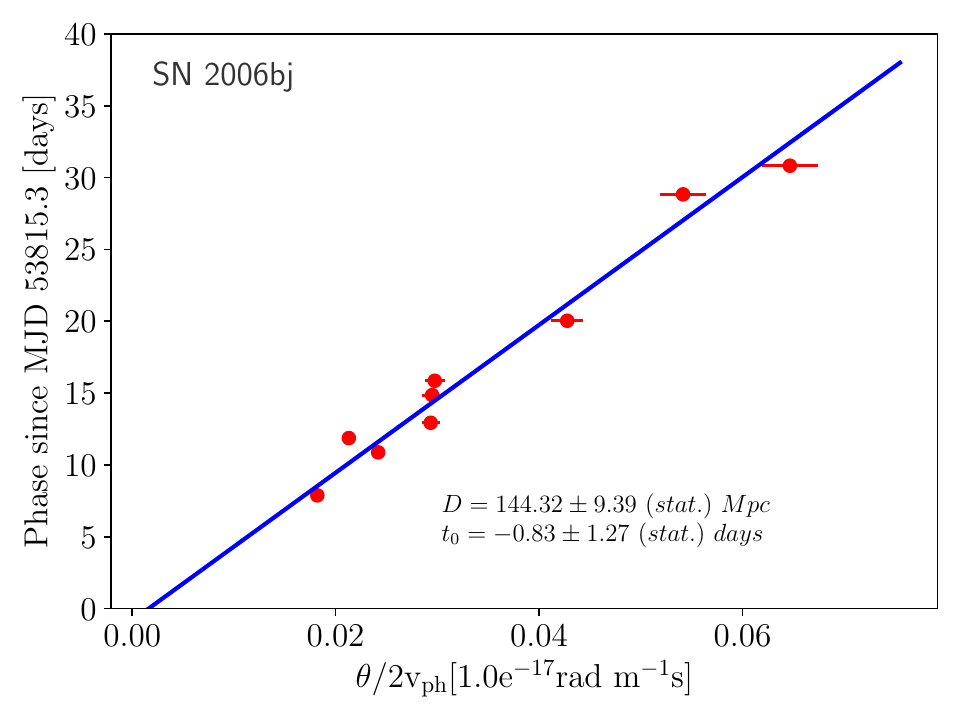}
\includegraphics[width=0.3\textwidth,height=0.2\textheight]{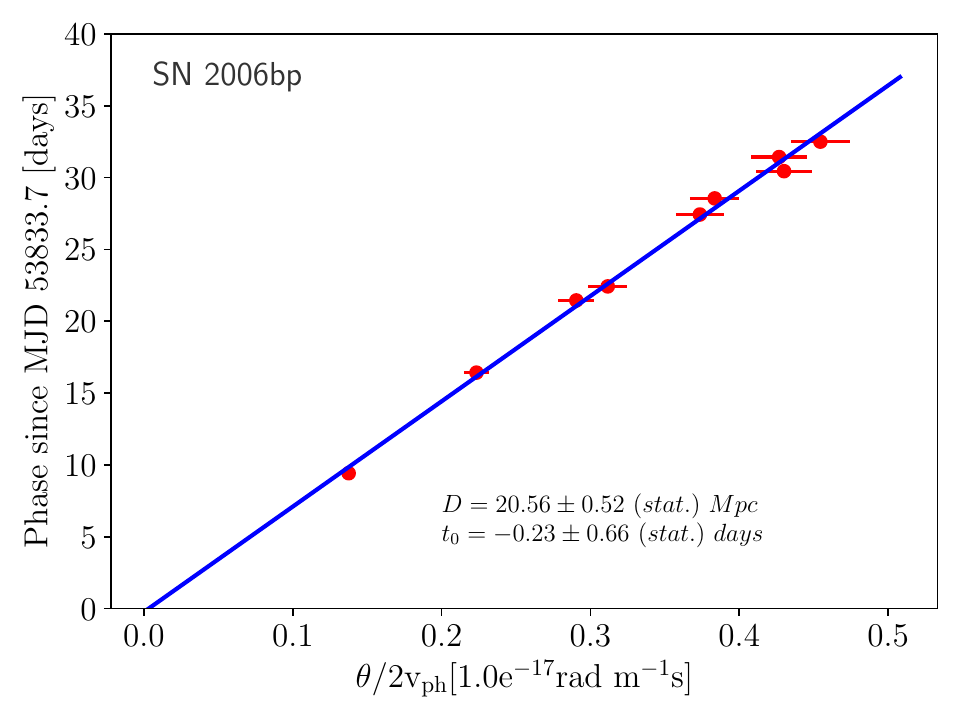}
\includegraphics[width=0.3\textwidth,height=0.2\textheight]{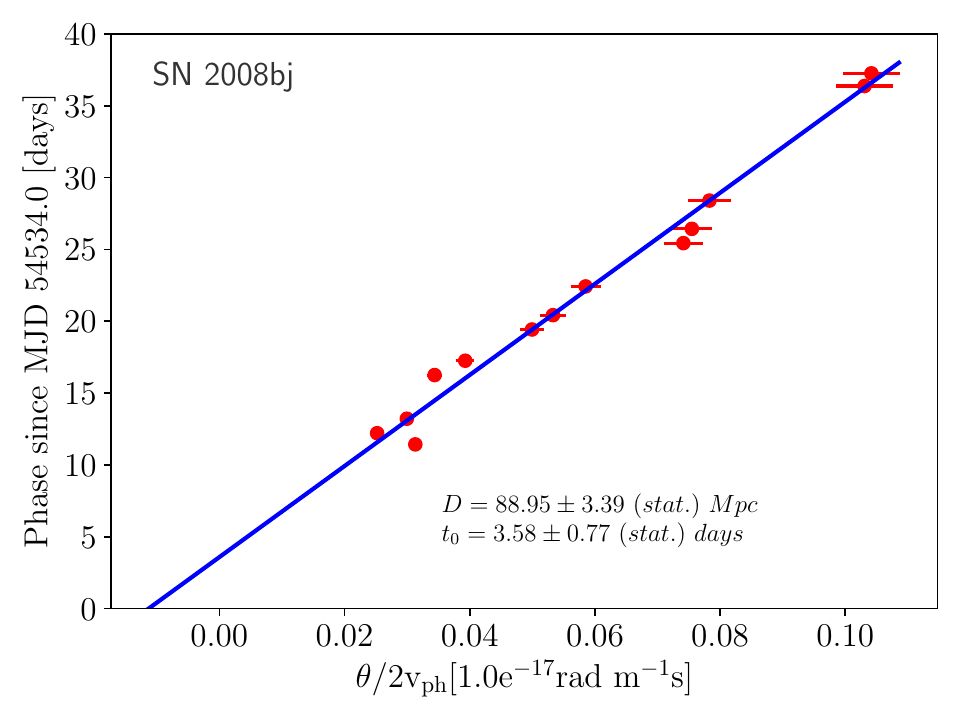}
\includegraphics[width=0.3\textwidth,height=0.2\textheight]{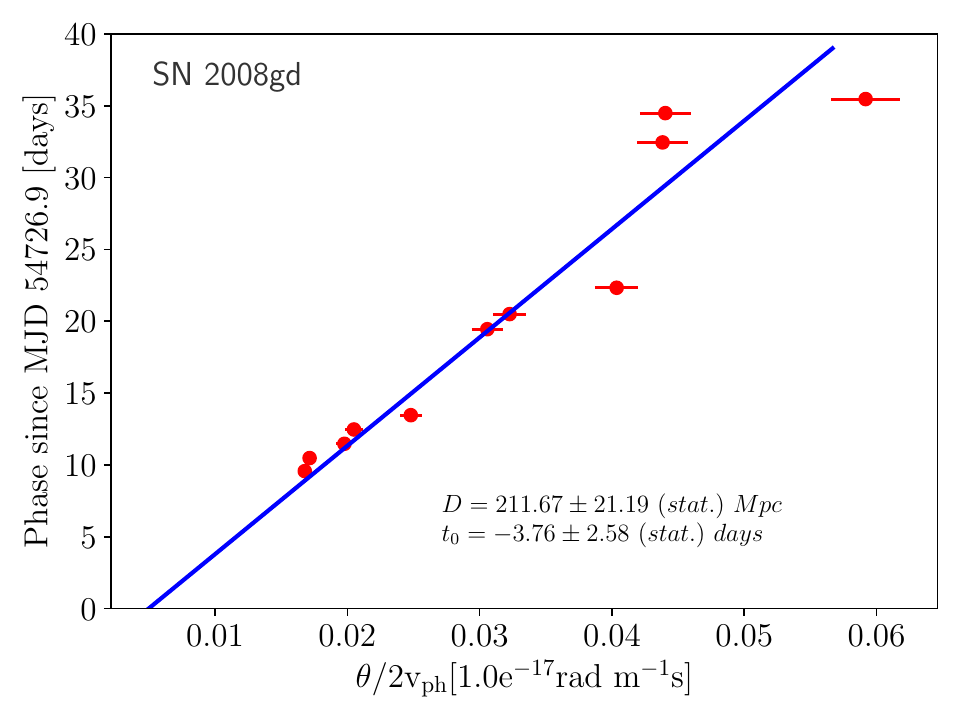}
\includegraphics[width=0.3\textwidth,height=0.2\textheight]{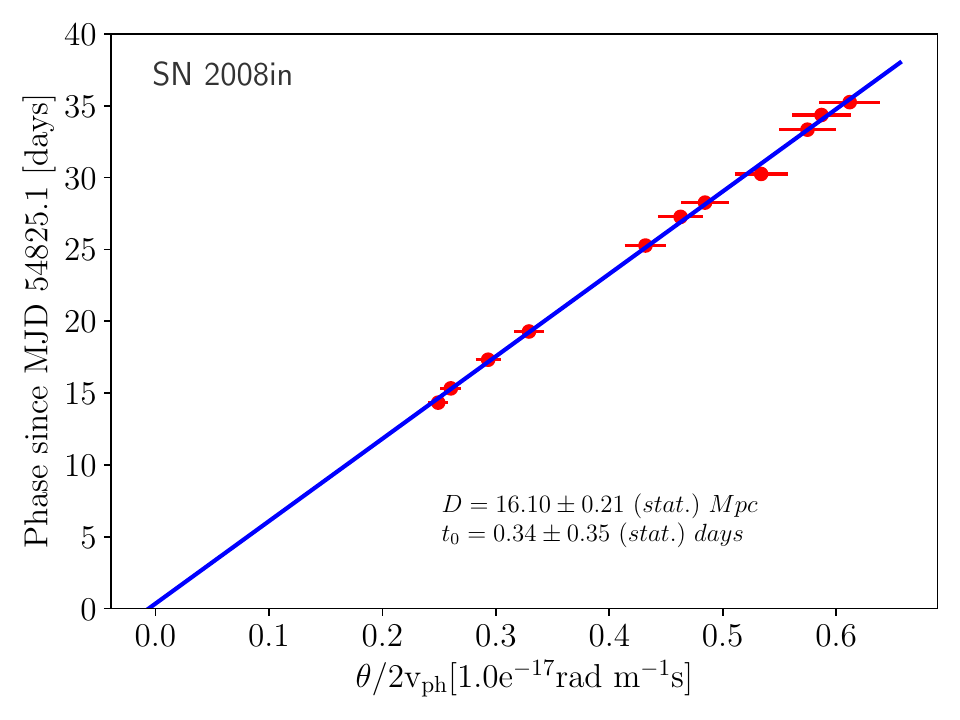}
\includegraphics[width=0.3\textwidth,height=0.2\textheight]{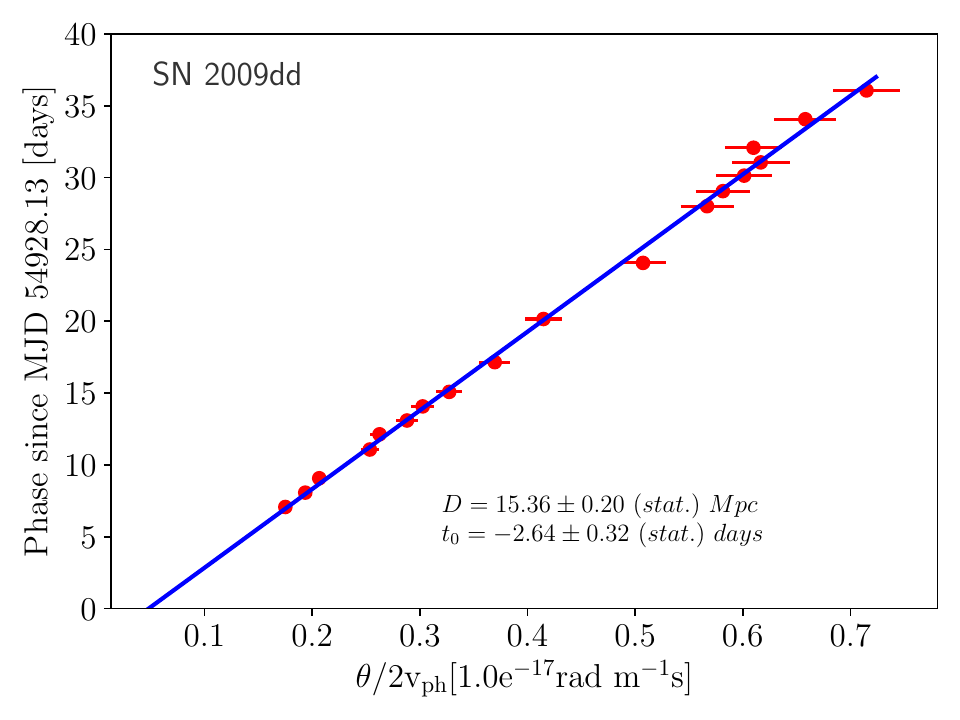}
\includegraphics[width=0.3\textwidth,height=0.2\textheight]{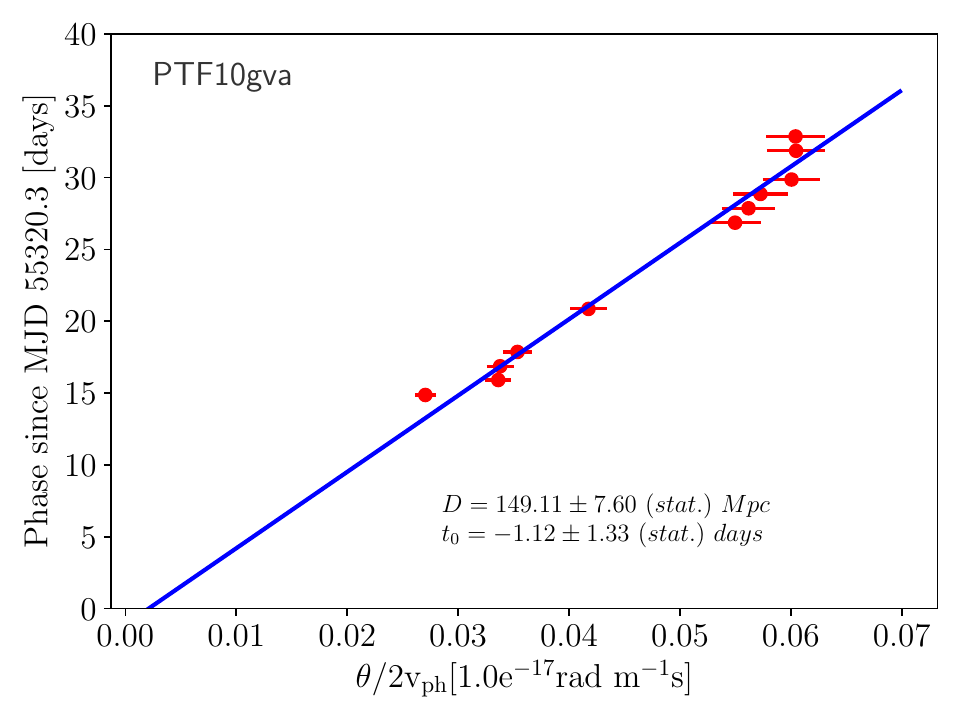}
\includegraphics[width=0.3\textwidth,height=0.2\textheight]{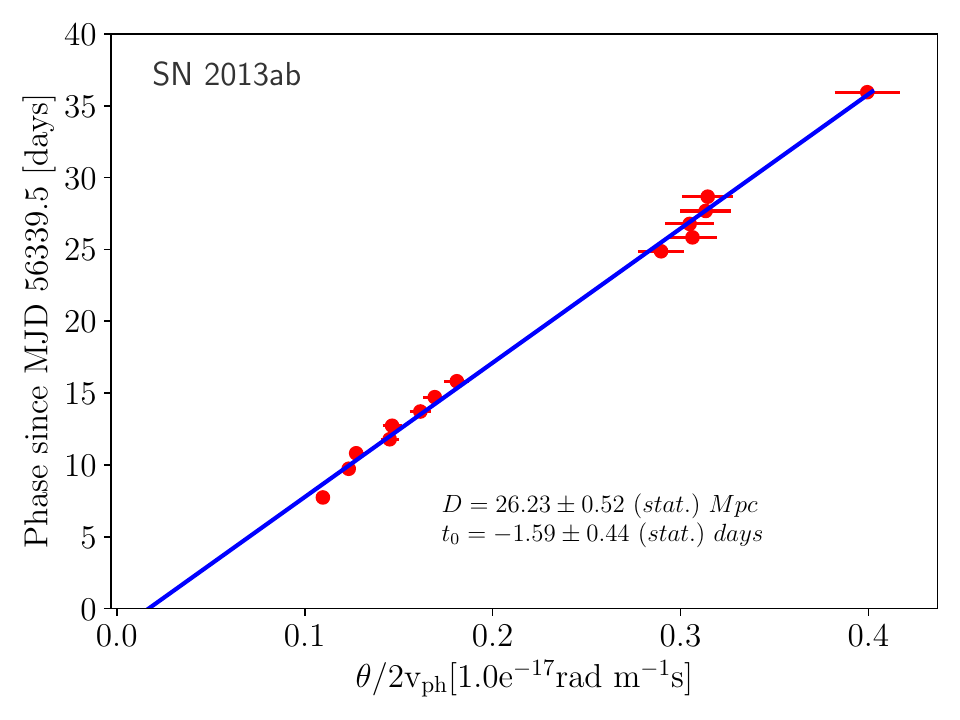}
\includegraphics[width=0.3\textwidth,height=0.2\textheight]{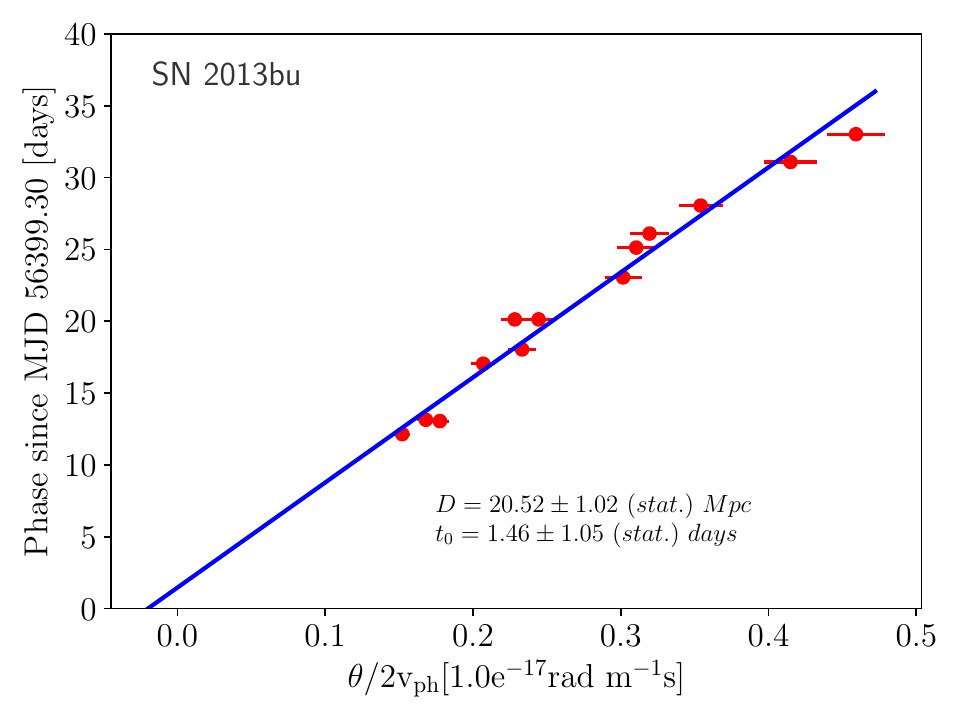}
\includegraphics[width=0.3\textwidth,height=0.2\textheight]{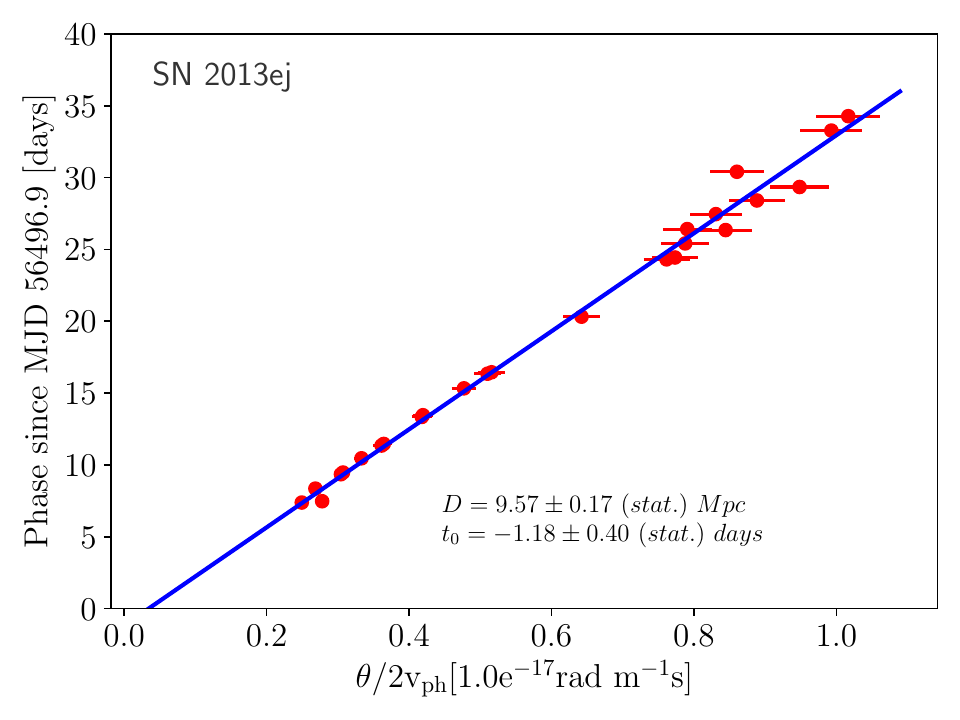}

\caption{Distance estimates of the 12 ROTSE SNe IIP using the EPM method.In each case of the SNe shown, the measured
data points are given at the ROTSE photometric epochs and the solid lines represent the best fit solution using Eq. \ref{eq:epmchi}.}
\label{fig:IIPsample_dist}
\end{center}
\end{figure*}

The systematic uncertainties in distance are also shown in Table \ref{tab:epmmeasure2}; where the total uncorrelated systematic uncertainties in distance are calculated from the $t_{0}$, E($B-V$), $z$ uncertainties and the uncorrelated uncertainties from the velocity and temperature evolution models, all added in quadrature. The total correlated systematic uncertainty includes the contributions from the velocity and temperature evolution models, added in quadrature.
Fig. \ref{fig:IIPsample_dist} shows the best fit estimated distance measured for each SN in our sample. 
In each of the SNe shown, the measured data points are given at the ROTSE photometric epochs and the solid lines represent 
the best fit solution using Eq. \ref{eq:epmchi}.

\begin{table*}
\begin{center}
\caption{Summary of the IIP parameters and Best fit EPM Distance and $t_{0}$}
\label{tab:epmmeasure2}
%\resizebox{\textwidth}{!}{\begin{tabular}{lcccc}
{\begin{tabular}{lccccccccc}
\hline
\hline
{SN} & $v_{\rm phot}(50)$  & $T$(50)  & \multicolumn{2}{c}{$t_{0}(days) $ } & \multicolumn{3}{c}{Distance($d$) (Mpc)} & & Median Distance from NED\footnote{Redshift independent distances from https://ned.ipac.caltech.edu/}\\
\hline
   &($10^{3}{\rm~km~s^{-1}}$) &($10^{3}$ K) &  &   & Fit & (stat.) & \multicolumn{2}{c}{(syst.)} & (Mpc)\\
\hline
   &  & & Fit    &    (stat.)      &     &         &  Total uncorr. & Total corr. & \\
\hline

  SN2004gy  & $4.5\pm0.2$ & $5.2\pm0.2$ & -3.88 &  1.36  & 115.78  &  6.39  & 12.04  &  6.06 & NA\\
  SN2005ay  & $3.9\pm0.1$ & $4.8\pm0.1$ & -0.17 &  0.55  &  22.33  &  0.54  &  3.62  &  1.14 & 21.9\\
  SN2006bj  & $4.4\pm0.2$ & $8.5\pm0.3$ & -0.83 &  1.27  & 144.32  &  9.39  & 24.07  &  7.77 & NA\\
  SN2006bp  & $4.3\pm0.1$ & $4.6\pm0.2$ & -0.23 &  0.66  &  20.56  &  0.52  &  1.76  &  1.07 & 17.5\\
  SN2008bj  & $5.2\pm0.2$ & $6.2\pm0.2$ &  3.58 &  0.77  &  88.95  &  3.39  & 11.62  &  4.36 & NA\\
  SN2008gd  & $4.9\pm0.2$ & $6.4\pm0.3$ & -3.76 &  2.58  & 211.67  & 21.19  & 32.17  & 10.96 & NA\\
  SN2008in  & $2.7\pm0.1$ & $5.5\pm0.2$ &  0.34 &  0.35  &  16.10  &  0.21  &  2.65  &  0.82 & 16.6\\
  SN2009dd  & $3.7\pm0.1$ & $4.6\pm0.2$ & -2.64 &  0.32  &  15.36  &  0.20  &  0.88  &  0.81 & 10.9\\
  PTF10gva  & $5.2\pm0.3$ & $6.5\pm0.2$ & -1.12 &  1.33  & 149.11  &  7.60  &  8.62  &  7.77 & NA\\
  SN2013ab  & $4.2\pm0.2$ & $4.9\pm0.2$ & -1.59 &  0.44  &  26.23  &  0.51  &  2.96  &  1.36 & 23.5\\
  SN2013bu  & $3.6\pm0.1$ & $3.7\pm0.2$ &  1.46 &  1.05  &  20.52  &  1.02  &  1.54  &  1.20 & 14.4(host)\\
  SN2013ej  & $4.3\pm0.1$ & $5.5\pm0.2$ & -1.18 &  0.40  &   9.57  &  0.17  &  0.29  &  0.49 & 9.5\\

\hline
\hline
\end{tabular}}
\begin{tablenotes}\footnotesize
The best fit distances and $t_0$ values for the 12 SNe IIP with respective statistical uncertainties from the EPM fits are given. Also shown are the total uncorrelated and correlated uncertainties for each of the distance measurements for the SN sample. In the rightmost column, median distances from NED catalog are also given for those SNe that have distances from redshift independent measurements. If no SN distance is available, host distances are given for reference whenever available.
\end{tablenotes}
\end{center}
\end{table*}

\section{Cosmological Analysis}\label{sec:cosmo}
 Using the luminosity distances derived above, we employ two methods to extract the Hubble constant ($H_0$) for the nearby universe.  In one approach, we simulate the impact of peculiar velocities on a sample of 12 galaxies with distances according to the observed distribution and fit for $H_0$.  We also perform a Markov Chain Monte Carlo (MCMC) to obtain a parametric estimation of $H_0$.  In our analysis, we assume no sensitivity of our measurements to other cosmological parameters such as $\Omega_M$ and $\Omega_{\Lambda}$. In each of these methods, we initially blind the fitted ($H_0$) parameter by translating with an unknown random additive scalar; which we un-blind at the end to obtain the final estimate.
 
\subsection{Linear Fit of $H_0$} \label{sec:linearfit}
Peculiar velocities complicate the extraction of $H_0$ from the sample and they are substantial relative to recessional velocity for the lower redshift constituents of this survey.  We attempt to mitigate this impact by simulating and fitting galaxies with peculiar velocities at the distances we have calculated.  Recessional velocities are simulated according to 
\begin{equation}
    \label{eq:vh0d}
    v= H_0 d
\end{equation}
where $H_0$ is a parameter that can be specified and we note that this linear relationship holds for the redshift range considered here.  We first simulate individual pseudo-experiments populated by SNe distributed according to the distances in Table ~\ref{tab:epmmeasure2}.  The recessional velocity for each galaxy in a pseudo-experiment is determined using Eq. \ref{eq:vh0d}. 

Peculiar motion is modeled for each simulated galaxy by adding a random velocity component from a Gaussian distribution centered around 300 ${\rm kms^{-1}}$, as is typically quoted in peculiar velocity studies (e.g. \cite{kessler09}, \cite{davis11}, \cite{johnson14}). Each generated peculiar velocity is multiplied by a random value between -1 and 1 to account for the component along the line of sight. Errors in distance for the simulated galaxies are chosen to reflect the data sample uncertainties. Errors in velocity are set to 300 ${\rm kms^{-1}}$. We split the analysis using a boundary at $cz_{CMB} = 3000 \rm kms^{-1}$, to distinguish whether peculiar velocity is large compared to the total velocity. Pseudo-experiments each including 7 galaxies with random distances of $5 \leq d \leq 30$ Mpc constitute the $local$ sample. Pseudo-experiments each including 5 galaxies with random distances of $70 \leq d \leq 260$ Mpc constitute the $H-flow$ sample.  We then fit Eq. \ref{eq:vh0d} to each pseudo-experiment, with $H_0$ as a free parameter forced to $0$ at $z=0$.

\begin{figure}
\begin{center}
\includegraphics[scale=0.3]{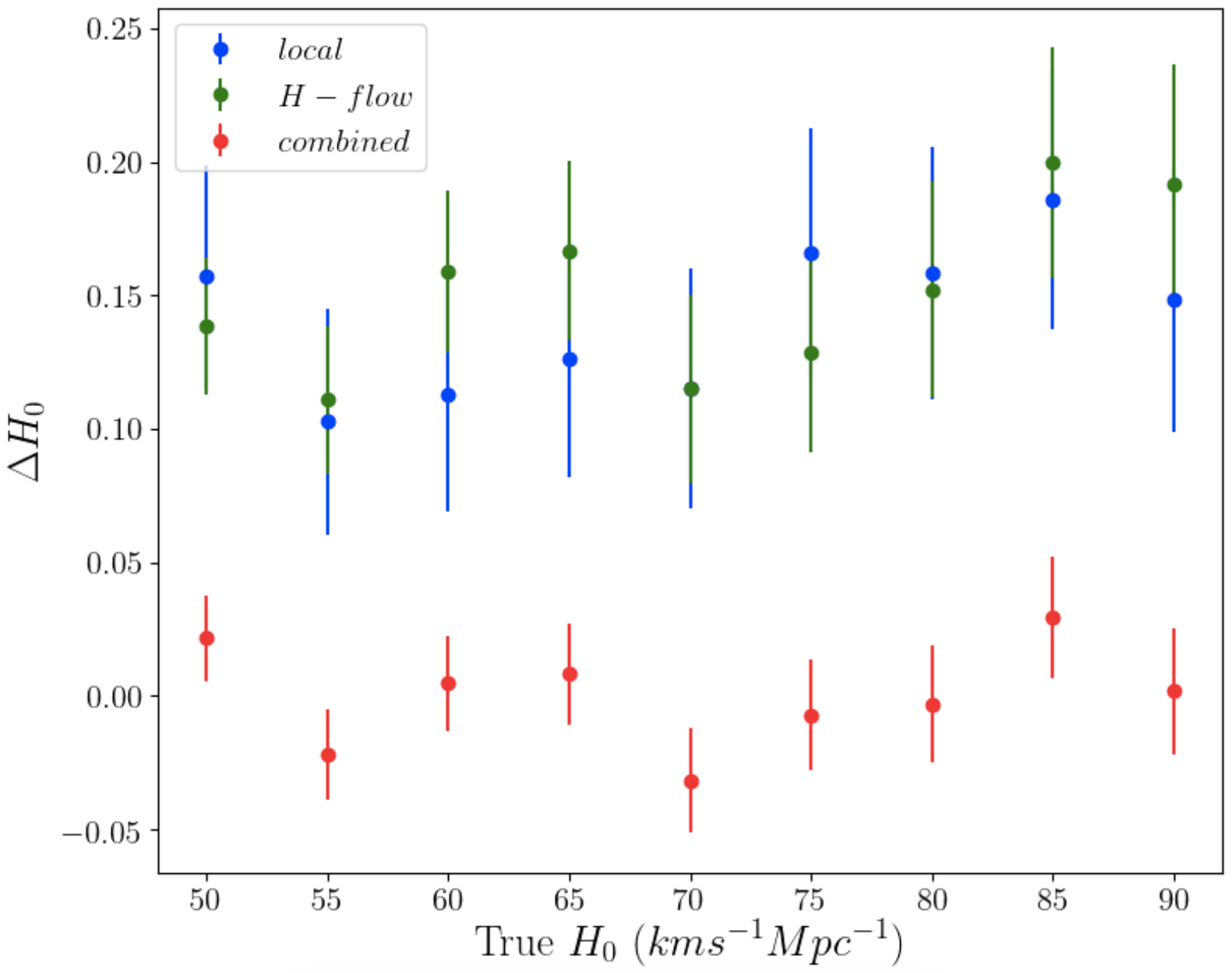}
\caption{Residuals of average $H_0^{meas}$ - $H_0^{true}$, $\Delta H_0$, for the simulated $local$ and $H-flow$ samples before calibration, and the $combined$ sample after calibration. See text for the slopes and offsets.
}
\label{fig:h0_calibration}
\end{center}
\end{figure}

We test this fitting approach for potential bias on the extracted $H_0$ measurement.  We generate 10,000 pseudo-experiments for both samples, repeating this process for input cosmologies of $50 \leq H{_0} \leq 90~{\rm kms^{-1}~Mpc^{-1}}$ at 5 ${\rm kms^{-1}~Mpc^{-1}}$ intervals. We calculate the measured value, $H_0^{meas}$, averaged over all pseudo-experiments of each ensemble with the specific input value, $H_0^{true}$. We apply a linear fit to $H_0^{meas}$ vs. $H_0^{true}$, and obtain the values of slope to be $1.0011 \pm 0.0007$ and intercept of $0.0651 \pm 0.0462$ for the $local$ sample. The corresponding fit parameters for the $H-flow$ sample are $1.0012 \pm 0.0007$ and $0.0669 \pm 0.0451$ respectively. These fits yield a slope and offset close to 1.0 and 0.0, respectively, for both the $local$ and $H-flow$ samples, indicating small but non-negligible bias. Using the fit parameters, we construct calibration vectors with values for each instance of input $H_0^{true}$. The overall $H_0^{meas}$ values for the ``$combined$'' sample are taken to be the weighted average of the $local$ and $H-flow$ sample measurements after the respective calibrations. A linear fit of $H_0^{meas}$ vs. $H_0^{true}$ for the $combined$ sample results in a slope and offset of $0.9999 \pm 0.0005$ and $0.0005 \pm 0.0371$, respectively, indicating an adequate mitigation. Differences between $H_0^{meas}$ and $H_0^{true}$ values, $\Delta H_0$, before and after calibration are shown in Fig. \ref{fig:h0_calibration} for a range of tested $H_0$ values.  The small estimated bias for each subsample and the efficacy of the corrections for the combined result, is independent of $H_0$.

 We apply the above procedure to the data, using the calibration vectors described above, for the $local$ and $H-flow$ samples, respectively. We obtain calibrated values of $H_0 = 67.4 \pm 11.8 (stat) ~{\rm kms^{-1}~Mpc^{-1}}$ for the $local$ sample and $H_0 = 75.6 \pm 5.1 (stat) ~{\rm kms^{-1}~Mpc^{-1}}$ for the $H-flow$ sample. The combined, weighted average yields a final value of $H_0 = 74.3 \pm 4.7 (stat)~{\rm kms^{-1}~Mpc^{-1}}$. The Hubble diagram from the fits of $local$, $H-flow$ and the final estimates are shown on the top panel of Fig. \ref{fig:hubble_diagram}. The blue and magenta dashed lines correspond to the fits to the $local$ and $H-flow$ samples and the red line correspond to the combined weighted estimate. 
 The vertical error bars include the total uncorrelated uncertainty shown from Table \ref{tab:epmmeasure2} added in quadrature with the estimated distance uncertainty from 300 ${\rm kms^{-1}}$ fixed peculiar velocity; while the horizontal error bars represent that peculiar velocity uncertainty. 
 
 We  assess an additional systematic uncertainty due to the correlated impact among the SNe from the velocity and temperature evolution model uncertainties.  We add up the correlated errors from Table~\ref{tab:epmmeasure2} for all SN distances and refit for $H_0$. We repeat shifting the SN distances low by the correlated errors.  The uncertainty on $H_0$ is calculated from the differences of these fits from the nominal $H_0$ value.  This results in a modeling uncertainty of $^{+4.1}_{-3.7}~{\rm kms^{-1}~Mpc^{-1}}$.  A systematic uncertainty due to the assumption of typical peculiar velocity of $300 {\rm kms^{-1}}$ was assessed by resimulating ensembles with $H_0 = 60, 70, 80 ~{\rm kms^{-1}~Mpc^{-1}}$ for distributions of peculiar velocity centered around 230 ${\rm kms^{-1}}$ and 370 ${\rm kms^{-1}}$ to estimate the lower and upper bounds of potential values of $H_0$ and peculiar velocity.  This yielded an uncertainty of $\pm 0.3 (sys)~{\rm kms^{-1}~Mpc^{-1}}$.

\subsection{Markov Chain Monte Carlo Sampling}

We also perform a parametric estimation of a cosmological model 
using an MCMC simulation in a Bayesian framework. For the given values 
of ($ z_{CMB}, \Omega_{M}, \Omega_{\Lambda}$), the luminosity distance ($D_{L}$) is given by

\begin{equation}
\label{eq:ld}
D_{L}=\frac{c}{H_{0}}\int_0^{z_{CMB}}{\frac{dx}{\sqrt{\Omega_{M}(1+x)^3+\Omega_{\Lambda}}}}
\end{equation}

\noindent We use the publicly available package EMCEE (\cite{foreemcee}), where the MCMC 
is performed to sample the posterior probability  distribution obtained from the given 
likelihood function and the distribution of the priors. We expand the 1-D likelihood function discussed previously in \citep{poznanski10,andrea10,deJaeger17} in 2 dimensional matrix form

\begin{equation}
\label{eq:likelihood}
\ln{\mathcal{L}} = -\frac{n}{2}\ln(det(V))-\frac{1}{2}\sum_{\rm i=1}^{n}(D_{i}^{ms} - {D_{L}}_{i})~V^{-1}~(D_{i}^{ms} - {D_{L}}_{i}) 
\end{equation}

\noindent where the sum is over ($n$=12) SNe in the sample. $D_{i}^{ms}$ is the measured distance using EPM and ${D_{L}}_{i}$ 
is the luminosity distance using Eq. \ref{eq:ld} for the $i^{\rm th}$ SN. The matrix $V$ is given as

\begin{equation}
\label{eq:covtot}
V = C + \sigma_{int}^2 I
\end{equation}

\noindent where $C$ is the covariance matrix and $\sigma_{int}$ is the residual intrinsic uncertainty that includes any unaccounted uncertainty in the analysis. The diagonal terms of the covariance matrix $C$ comprise the statistical uncertainties added in quadrature 
with the total uncorrelated systematic uncertainties from the Table \ref{tab:epmmeasure2}, while the off-diagonal terms 
are computed by multiplying the total correlated systematic errors for respective SNe from Table \ref{tab:epmmeasure2}. 
As in the linear fit analysis, we add in quadrature an uncorrelated uncertainty of 300 ${\rm kms^{-1}}$ in the diagonal terms to account for the contribution due to peculiar motion. 

We consider a flat universe ($\Omega_M+\Omega_\Lambda =1; \Omega_M = 0.3$) as our prior. The only free parameters are the 
Hubble parameter ($H_{0}$) and the intrinsic uncertainty $\sigma_{int}$. Now we are not only interested in the best fit values of 
these parameters, but also in their full maximum $a~posteriori$  probability density distribution. In the Bayesian 
framework, the joint posterior probability function for these parameters can be written as 
\begin{multline}
p(H_{0},\sigma_{int}|z_{CMB},D_{L},C,\Omega_{M},\Omega_{\Lambda}) \\
~~~~~~~~~~\propto p(H_{0},\sigma_{int})p(D_{L}|z_{CMB},C, H_{0},\sigma_{int},\Omega_{M},\Omega_{\Lambda})
\end{multline}
\noindent The function $p(D_{L}|z_{CMB},C, H_{0},\sigma_{int},\Omega_{M},\Omega_{\Lambda})$ is the likelihood 
function $\mathcal{L}$ given in Eq. \ref{eq:likelihood}; while for the prior distribution, $p(H_{0},\sigma_{int})$, we chose flat priors given by
\begin{eqnarray}
50 {\rm ~kms^{-1}~Mpc^{-1}} ~<~H_{0}~<~150~ {\rm kms^{-1}~Mpc^{-1}} \\
0~{\rm Mpc}~<~\sigma_{int}~<~50~{\rm Mpc}
\end{eqnarray}

\noindent We first evaluate the best fit maximum likelihood estimation (MLE) values of $H_{0}$ 
and $\sigma_{int}$ by minimizing the negative log of the likelihood in Eq. \ref{eq:likelihood}. 
Next, we initialize the MCMC chains by picking 500 random initial points  as seeds by sampling a small 
2-D Gaussian ball about the MLE of the parameters $H_{0}$ and $\sigma_{int}$. The MCMC is performed using the EMCEE framework for 
500 steps for each walker; resulting in a joint posterior probability distribution $p(H_{0},\sigma_{int})$. 
To avoid any systematic sampling bias from the choice of initialization, we discard the first 50 steps of each walker; 
after which the walkers begin to fully span the full posterior distribution. 

\begin{figure}
\includegraphics[width=0.45\textwidth]{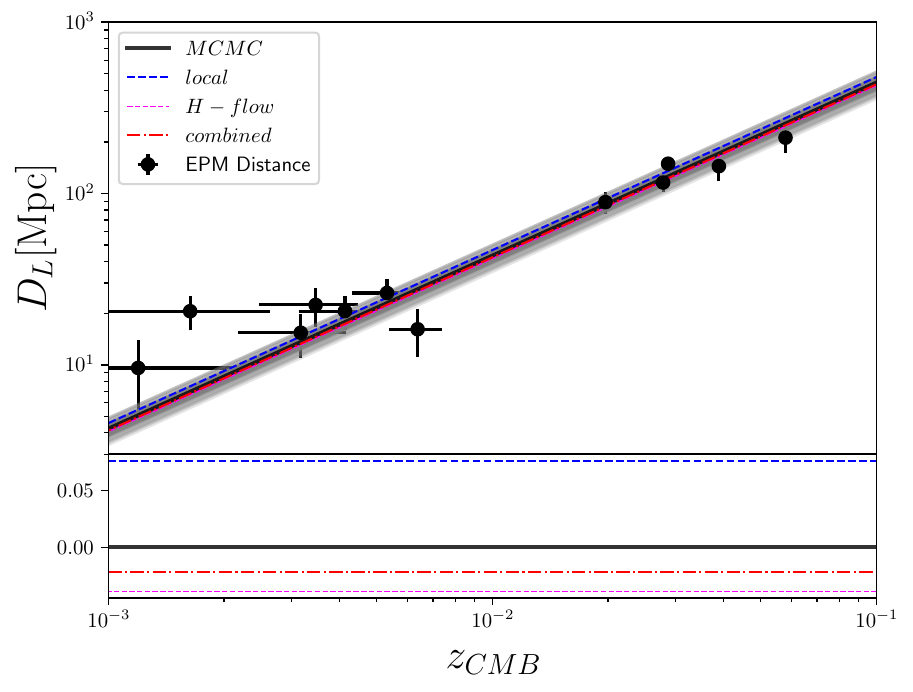}
\caption{Top: Hubble diagram showing the posterior samples from MCMC in gray.
The solid black line is the maximum likelihood estimate from the MCMC method. The black points are the measured luminosity distances using EPM. The blue and magenta dashed lines correspond to the linear fits to the $local$ and $H-flow$ samples and the red curve is the combined weighted estimate from the linear analysis. The horizontal error bars on the data points correspond to the assumed peculiar velocity of 300 ${\rm kms^{-1}}$. Bottom: Residuals of the linear fits with respect to the best fit estimate from the MCMC analysis, shown as difference of log values}
\label{fig:hubble_diagram}
\end{figure}

The resulting Hubble diagram from the posterior samples 
are shown in the top panel of Fig. \ref{fig:hubble_diagram}. The solid black line represents our best fit model from the MCMC analysis; and the gray lines represent 
the posterior samples from the MCMC runs. The black data points are the measured luminosity EPM distances for the SNe IIP sample. 
The vertical and horizontal error bars are described in Section \ref{sec:linearfit}. From the final marginalized 1-D posterior distributions, the Hubble parameter is estimated to be $72.9^{5.7}_{-4.3} ~{\rm kms^{-1}~Mpc^{-1}}$ and the intrinsic scatter $\sigma_{int}$ is estimated to be $0.3^{0.42}_{-0.26} {\rm Mpc}$.  
The bottom panel of Fig. \ref{fig:hubble_diagram} shows the residuals of the obtained fits from the linear analysis in Section \ref{sec:linearfit} relative to the best fit from the MCMC analysis shown in the difference of log values.

\section{Results and Discussion} \label{sec:discuss}

Our aim has included development of a methodology for distance measurement of SNe IIP utilizing unfiltered CCD photometry and minimal color and spectroscopic data, and that would enable a sensitive probe of cosmic expansion.  We had addressed this previously for SN 2013ej \citep{dhungana16}, which was far from the host nucleus and did not require an image differencing approach.  However, analysis of a more representative survey of SNe IIP required a more general approach.  

The new image differencing algorithm, $\texttt{ImageDiff}$, introduced in this paper was designed to obtain photometric measurements with high efficiency and consistency over a wide range of host environments. It also delivered photon-limited measurements unimpacted by artifacts and their attendant systematic uncertainties.  The performance was assessed by injecting PSFs into an image with a full range of SN-nucleus displacements.  This showed that we could extract the simulated magnitudes precisely up to a limiting magnitude of $\sim18$ and across the wide field of view. The performance study showed no significant bias in the photometric residuals as shown in Fig. \ref{fig:diffsim}. The new software demonstrated improved photometric detection efficiency up to ∼20\%.  By examining SN 2004gk, a typical SN with distance to host nucleus of $\sim3''$, we are able to quantify the improvement over the prior image differencing code for ROTSE.  The scatter in points around the known lightcurve is reduced by approximately 3 times to roughly 0.1 magnitude (Section \ref{sec:imagediffperf}) when compared to prior approaches used for the ROTSE SNe data reduction pipeline. The increased performance was particularly observed in the crowded fields and when the SN was closer to the host nucleus. We also note that for similar image sizes in most cases, the performance across several kernel bases were similar.  These results are critical to what follows because the astrophysical and distance measurements produced in this study rely on minimizing photometric uncertainties overall.  
 
In \cite{dhungana16}, we demonstrated with SN 2013ej the ability to calibrate well the ROTSE unfiltered photometry to a pseudo-bolometric magnitude to facilitate the EPM extraction of distance.  Such a method requires substantial color photometry spanning from the near ultraviolet to near infrared wavelengths at various epochs of the supernova.  While that method yielded excellent photometry and an accurate distance measurement, we have attempted a different approach here that relied less on such extensive filtered photometry and spectroscopy by calibrating the ROTSE magnitude to the $V$-band magnitude and using color-based temperature estimates to improve the correction.  We also established a V band calibration for the ROTSE unfiltered SNe lightcurves that were consistent with the observed V band SNe lightcurves from the literature.  In the end, the correlation with an actual V-band magnitude, at 0.01 magnitudes, is of sufficient precision to obtain accurate distances while keeping the extensiveness of the required broadband photometry and spectroscopy to a minimum. 
 
Utilizing filtered photometry and spectroscopy for several well measured SNe IIP, we saw that the velocity and temperature evolution exhibited similar behaviors. To quantify this evolution and provide a model from which these properties can be determined at any epoch, we established empirical calibrations for both properties' evolution using single epoch photometric and spectroscopic measurements (Section \ref{sec:prop}). The exponential decay behavior has been empirically observed in the literature of SNe IIP, but its physical origin appears to require further study. Modeling of the photospheric velocity evolution from the spectra indicates that the use of Fe II lines only, such as in \cite{Faran14}, does not accurately describe the early behavior when extrapolated.  Using Fe II when observed and He II and H-alpha lines during early times makes the decline in velocity less steep in agreement with observed SNe (Section \ref{sec:vphotmodel}). This model also seems to improve the agreement with data sufficiently late in the plateau.  Velocities measured for our IIP sample ranged from 2700 ${\rm kms^{-1}}$ to 5200 ${\rm kms^{-1}}$ at $t~=~50$d.  Remarkably, when normalizing values from other epochs to this value, the behavior for different events present a very precise exponential fall off with $t$; supporting the ability to calibrate the evolution from measurement at a single epoch. Interestingly, the temperature evolution appears to be well described by a similar exponential profile.  The temperatures at 50d range widely from 3700 to 8500 K and yet the evolution lines up well, with the exception of SN 2005cs which is slightly steeper at early times and flatter than other SNe later. SN 2005cs has been very well studied in the literature and appears to be an outlier from the general population as an underluminous SN. Nevertheless, the four SNe we examined yielded a good fit to Eq. \ref{eq:tempIIp}.   

Use of the EPM technique for distance measurements provided additional constraints on the supernova itself via 
the fit to $t_0$.  The precision yielded was as small as $\pm 8$ hours. The fitted $t_0$ from the EPM were found consistent with our initially adopted values of $t_0$. We were able to calculate the luminosity
distance with limited unfiltered photometry and spectroscopy.  Distances ranged from $9.57\pm0.17({\rm stat.})\pm0.29({\rm uncorr.})\pm0.49({\rm corr.})$ to $211.67\pm21.19({\rm stat.})\pm32.17({\rm uncorr.})\pm10.96({\rm corr.})~{\rm Mpc}$.  As shown in Table \ref{tab:epmmeasure2}, 
these distance measurements agree with those in the literature from the host galaxies.  

The precise calibration of astrophysical properties using a single measurement epoch, as in 
several of these SNe, can be a powerful advantage of the EPM over other techniques, where the observations need not only to be densely sampled but also require concurrent multi-band photometry.  Given the potential challenges
of observing with large pitch, unfiltered CCDs, the measurements in Fig. \ref{fig:IIPsample_dist} are remarkably linear thru the entire
range of epochs chosen for all of the SNe in this sample.  Dense sampling during the plateau phase increases
statistics and could reduce the EPM uncertainties.  However, the low scatter of points around these slopes, which
arises in part from the low scatter of points around the physical parameter evolution models, suggests observables
have strong correlations and numerous concurrent observations are not absolutely necessary for the EPM measurements.  

These results strongly suggest a robustness of the method to the choice of plateau span utilized. In particular, stripped core supernovae of types Ib and
Ic that lack extensive plateaus might be viable in this method using a shorter time duration.  Such a possibility was already explored in \cite{vinko04} with SN 2002ap.  We have
also obtained a preliminary result for 2007gr in \cite{statenPhd}.  Both  yielded distances to their hosts
in agreement with the literature. 

\begin{figure}
\begin{center}
\includegraphics[width=0.5\textwidth]{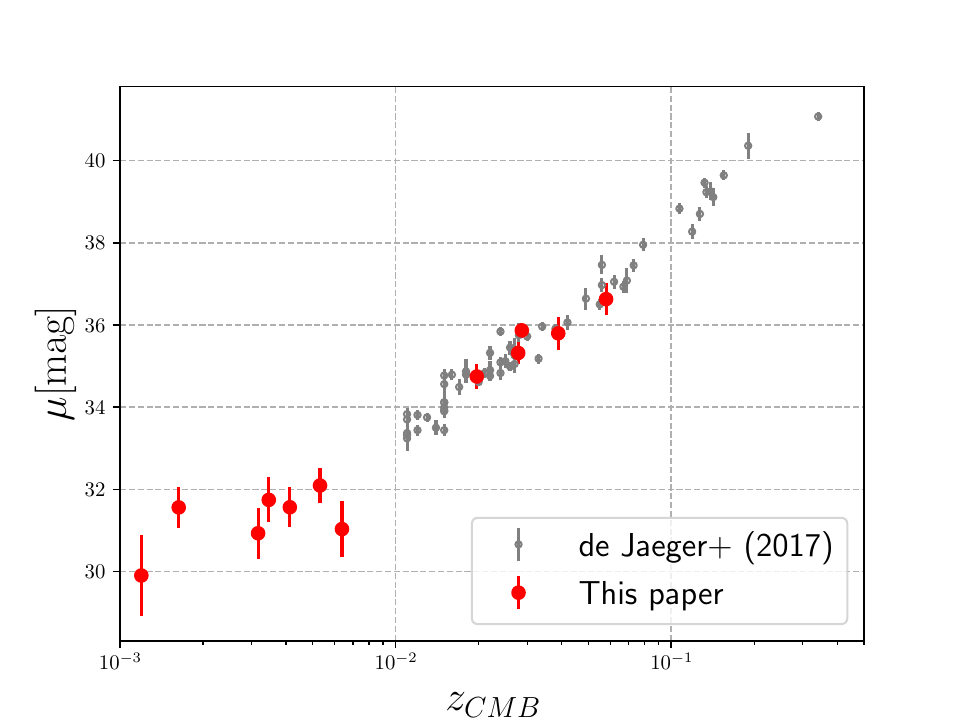}
    \caption{Distance moduli for the 12 SNe in our sample using the EPM method are overlaid on top of the SCM measurements from \cite{deJaeger17}. The distance uncertainties for our sample include the contribution from flat peculiar velocity of 300 $kms^{-1}$ along with the statistical and total uncorrelated uncertainties from Table \ref{tab:epmmeasure2}, added in quadrature.}
    \label{fig:comparedist}
    \end{center}
\end{figure}

Our analysis shows an inherent promise of the EPM method whose very general physical assumptions have allowed us
to leverage unfiltered photometry with minimal spectroscopy and color information.  In fact, the results indicate
that SNe with only one epoch of either performed similarly to those with 3 or more epochs when using the
normalized, exponential time evolution models described in Section \ref{sec:prop}.  Our results can also be obtained by using V
band photometry in place of the unfiltered photometry.  We note that while the accurate systematics from the
dilution parameter may affect the distance results, some limitation in their magnitude can be inferred from the
linearity of the EPM plots over the whole plateau, and by the validity of the SNe IIP distances compared
to their host's.  Further work, where the uncertainties will be reduced overall, will necessitate more careful handling of this issue. Fig \ref{fig:comparedist} shows the EPM distance modulus derived for our SN sample overlaid on top of those measured from the SCM technique by \cite{deJaeger17} (see Fig. 5 in their paper). On the overlapping redshift range, the distances measured from independent methods and samples are statistically consistent.

A driving motivation in the current study was to test cosmic expansion in the local universe.  The linear method we used has the advantage that it makes no cosmological assumption and allowed us to estimate and mitigate the effects from peculiar motion via a simple simulation.  As such, it yielded consistent values of $H_0 =67.4\pm11.8({\rm stat.})~{\rm kms^{-1}~Mpc^{-1}}$ and $H_0= 75.6\pm5.1({\rm stat.})\pm0.3({\rm syst.})~{\rm kms^{-1}~Mpc^{-1}}$ in the $local$ and $H-flow$ regimes, respectively, with very different relative sizes of peculiar velocities to cosmic recession.  The combined measurement yields $H_0= 74.3^{+4.7}_{-4.7}({\rm stat.})^{+4.1}_{-3.7}({\rm syst.})\pm0.3 {\rm ~(Pec.~Vel.)~kms^{-1}~Mpc^{-1}}$.  The impact of peculiar velocities is minimal.  We can also see that with only 12 SNe, we are already almost systematics dominated.  Future work will require more effort to reduce systematic effects.

The MCMC method we employed presented different advantages to the analysis.  The technique jointly estimates the posterior distribution for $H_0$ and a measure of intrinsic residual systematic uncertainty $\sigma_{int}$ for the analysis.  We utilize the full information in the covariance matrix, and the approach provides a test of unknown contributions to the scatter of points.  The value returned, corresponding to $0.3^{+0.42}_{-0.26} Mpc$ is consistent with zero.  We obtain a measurement for the full sample of 12 SNe of $H_0= 72.9^{+5.7}_{-4.3} ~{\rm kms^{-1}~Mpc^{-1}}$, which is in excellent agreement with the result from the linear method.  Even though we have fixed other cosmological parameters, it is interesting to note that the residual intrinsic dispersion we find from the analysis of our SNe IIP sample is not statistically significant. 

For a quick test, we repeated the MCMC analysis using the 5 events from the $z > 0.01$ sample but dropping the peculiar velocity from the uncorrelated diagonal error term. This choice of cutoff is ubiquitous in the literature including the SNe Ia studies. The contribution from the galactic peculiar motion at such redshift is expected to be $\sim$10\% or less (e.g. \cite{kessler09}). In such a limited sample, we observe no statistically significant shift in either $H_0$ or $\sigma_{int}$.  Because we incorporated the correlated systematic uncertainties within the covariance matrix of the distances in the MCMC analysis, we take this estimate as our final estimation for $H_0$. 

When comparing to the other probes such as SNe Ia measurements from recent DES results (\cite{abbott19}), we see that our results are in good agreement. We also find our estimated $H_0$ value to be consistent with that obtained from the CMB measurements (\cite{planck16}) at the $1\sigma$ level.  In the future, our measurement can be substantially improved by a larger sample, and by more careful consideration of the underlying time evolution modeling and correlations among SNe we have utilized.

\section{Conclusion} \label{sec:conclusion}
We performed an end to end analysis of time evolution of SNe IIP properties and of cosmological properties measurement in the EPM framework using photometric and spectroscopic observations of a sample of 12 SNe IIP.  We 
significantly improved the ROTSE SNe photometry sampling and precision with new image differencing 
for ROTSE SNe images using a kernel convolution technique.  

In our analysis of SNe IIP, we 
have established excellent performance of unfiltered CCD photometry, including in areas crowded by host 
nuclei, to yield valuable measurements of supernova properties and to measure cosmic expansion in the 
nearby universe.  We demonstrated a broad consistency between SNe IIP of the time evolution of event 
ejecta velocities and photospheric temperatures from the times of peak luminosity throughout most of the plateau among these diverse SNe.  We empirically established parametric evolution models to extrapolate 
the photospheric velocity and temperature from as few as a single photometric and spectroscopic 
measurement.  Using the EPM technique, we obtained the luminosity distances 
for each SN. These distance measurements are in good agreement with host distances in the literature, 
and the linearity of the EPM diagrams suggests the viability of further generalization of this approach.  
Overall, the EPM technique looks promising to pursue cosmological studies for larger data sets, potentially 
to even higher red shifts.  

Utilizing two approaches to fitting for $H_0$, we have obtained a measurement 
of $H_0= 72.9^{+5.7}_{-4.3} ~{\rm kms^{-1}~Mpc^{-1}}$.  We further established that unknown peculiar velocities do not significantly impact this measurement.  Results from an MCMC approach also indicate that we have accounted 
for all appreciable contributors, and their uncertainties to the scatter of points.

\section{Acknowledgement} \label{acknow}
RK wishes to thank NASA grant NNX10A196H (P.I. Kehoe), and SMU’s Dean’s Research Council, for supporting the initial work for this paper. JCW and JV are supported in part by NSF AST-1813825. JV and his group at Konkoly Observatory are supported by the project ``Transient Astrophysical Objects'' GINOP 2.3.2-15-2016-00033 of the National Research, Development and Innovation Office (NKFIH), Hungary, funded by the European Union.
We would like to acknowledge the McDonald Observatory technical and observation support crew for tremendous help during the ROTSE and HET observations over the years. We would like to thank Jeffrey Silverman for useful discussions on the data processing and analysis.

\newpage
\newpage
\end{document}